\newtheorem{theorem}{Theorem}
\newtheorem{proposition}{Proposition}

\newtheorem{lemma}{Lemma}

\documentclass[12pt]{article}

\usepackage{amsmath}
\usepackage{amssymb}
\usepackage{amsfonts}

\newcommand{\qed}{{\mbox{} \hspace*{\fill}{\vrule height5pt width4pt
depth0pt}}\\}

\def\M{\hspace*{0.75em}}

\begin{document}

\title{Analysis of Markovian Competitive Situations using Nonatomic Games}
\author{Jian Yang\\
Department of Management Science and Information Systems\\
Business School, Rutgers University\\
Newark, NJ 07102\\
Email: jyang@business.rutgers.edu}
%Phone: 973-596-3658}

\date{
	%June 2012; revised,
	%August 2015
	%June 2016
November 2016; revised, February 2017}
\maketitle

\begin{abstract}
For dynamic situations where the evolution of a player's state is influenced by his own action as well as other players' states and actions, we show that equilibria derived for nonatomic games (NGs) can be used by their large finite counterparts to achieve near-equilibrium performances. We focus on the case with quite general spaces but also with independently generated shocks driving random actions and state transitions. The NG equilibria we consider are random state-to-action maps that pay no attention to  players' external environments. They are adoptable by a variety of real situations where awareness of other players' states can be anywhere between full and non-existent. Transient results here also form the basis of a link between an NG's stationary equilibrium (SE) and good stationary profiles for large finite games.\vspace*{.1in} %Our approach works well for certain dynamic pricing games, both without and with production.\\


%\noindent{\bf MSC2000 Classification Codes: }60G50; 91A10; 91A13

%\noindent{\bf JEL Code: }C72--Noncooperative Games

\end{abstract}

\newpage

\section{Introduction}\label{introduction}

%\subsection{Description of Research}

Many multi-period competitive situations, as first noted by Shapley \cite{S53}, involve randomly-evolving player states that affect players' payoffs. When making a decision, a player has to contemplate not only what states other players are in and how other players will act, % as in game-theoretic studies,
 but also how his and others' states and actions will influence the future evolution of all  players' states. % as in Markov decision processes.
%The state transition of each player from a period to another is also influenced by the idiosyncratic (local) shock experienced by the current player.
Another complicating factor is that players may have zero, partial, or full knowledge of other players' states before they take their actions in each period. The task of analyzing these dynamic games is certainly daunting. Consider a dynamic pricing game as an example. Multiple firms start a fixed time horizon with stocks of the same product. Each of them is bent on using pricing to influence demand and earn the highest revenue from selling their respective stocks. In any given period, a firm is aware of its own current inventory level but not the levels of others. Yet, demand arrival to the firm is both random and influenced by not only its own price, but also prices charged by other firms.
%In addition, the actual demand level is susceptible to an idiosyncratic shock that is independently generated in every period for every firm.
%In spite of its practical relevance, this dynamic pricing game has not received satisfactory treatment due probably to its obvious complexity.

The ultimate goal with such a Markovian game lies in identifying an equilibrium action plan that will earn each player the highest total payoff when other players adhere to the plan. But even in the stationary setting, known equilibria come in quite complicated forms that for real implementation, demand a high degree of coordination among players; see, e.g., Mertens and Parthasrathy \cite{MP87}, Duffie, Geanakoplos, Mas-Colell, and McLennan \cite{DGMM94}, and Solan \cite{S98}. Alternatively, we propose that equilibria be reached asymptotically as the number of players grows, on the premise that the game's nonatomic-game (NG) counterpart is analyzable. In the latter, a continuum of players are in competition, none of whom having any discernible influence on any other players and yet all players in aggregation hold sway on players' payoffs and state evolutions. The key advantage of such a game is that its state distribution will evolve in a deterministic fashion. This results in the relatively simple form taken by the NG's equilibria $x$: the pure or mixed action plan $x_t(s_t)$, though dependent on the time period $t$ and his own individual state $s_t$, is insensitive to whatever portion of the overall state distribution that the player can observe. %Typically, there is more chance for the NG to be analyzable than the original multi-player system.

%This being said, we caution that the state-distribution evolution pathway say $\sigma=(\sigma_1,\sigma_2,...)$ still implicitly affects players' action choices in an NG whose initial state distribution is given as $\sigma_1$. It can be said that the totally predictable $\sigma$ is on the back of every player's mind. An action plan $x$ is the NG's equilibrium just because it is an optimal response to a certain $\sigma=(\sigma_1,...)$; at the same time, the initial distribution being $\sigma_1$ and the fact that every player plays $x$ will together lead to the $(\sigma_2,\sigma_3,...)$ portion of the said $\sigma$. For our NG with Markovian feature, the environment-to-decision direction from a given $\sigma$ to its optimal response $x$ is itself dynamic programming in nature, due to a player's randomly evolving individual state $s_t$. Still, the NG is easier to handle than a finite-player counterpart or one with aggregate shocks. In any of the latter, the pathway $\sigma$ can no longer be taken to be deterministic. An action plan has to be in the form of $x_t(s_t,o_t)$ rather than $x_t(s_t)$, where $o_t=\tilde o(\sigma_t)$ represents a player's real-time observation of his period-$t$ surrounding $\sigma_t$. In finite games, there is the added difficulty of different players experiencing different external state distributions.

When an NG equilibrium is handy, we show that it can be used on the original finite Markovian game to serve our intended purpose. Relying on intermediate results stemming from the weak Law of Large Numbers (LLN) concerning empirical distributions,
%most notably Lemmas~\ref{joint}, ~\ref{cojoint}, and~\ref{T-onestep},
we establish two main results. In Theorem~\ref{T-converge}, we show that the empirical distribution of players' states, which is itself random in the finite game, will nevertheless converge in probability to the deterministic distribution as predicted for the NG counterpart when the number of players grows to infinity. This convergence paves way for Theorem~\ref{main}, which states that players can apply the observation-blind NG equilibrium to the finite-player situation and gain an average performance that is ever harder to beat as the number of players grows. In both results, the ``average'' on players' states is assessed on the state distribution prevailing in either the NG or the finite game. After assuming time-invariant payoff and transition functions, as well as fixed discountings over time and an infinite time horizon, we obtain a stationary setting. For this, we establish Theorem~\ref{main-s}, effectively our affirmative answer to whether stationary equilibria (SE) studied in past literature can be useful in large finite games. %The derivation of this result through the transient Theorem~\ref{main} indicates that transient results are probably more fundamental.  %It is also not difficult to introduce exogenous global states to our setting. %  too. These states can be used to model commonly felt economic/environmental conditions like the inflation-adjusted Brent Price Index or Dow Jones Industrial Average. With their presence, the state-distribution pathway even in an NG is no longer deterministic. Rather, it is under the sway of the global-state history (GSH). Theorem~\ref{main-g}, the GSH-dependent version of Theorem~\ref{main}, conveys essentially the same message about the suitability of NG equilibria to large finite games.

The above theory will be most useful when the NG counterpart is relatively easy to deal with in comparison to the corresponding finite games. Besides evidence in literature, this point is further buttressed by the dynamic pricing game mentioned earlier. Presented in Yang \cite{Y17} as supplementary material to the current paper, our analysis demonstrates the usefulness of the transient result Theorem~\ref{main}. The game is also extended through the consideration of locked-in production, wherein every firm uses production to bring its inventory back up to a pre-determined level whenever it becomes empty. The resultant game is again asymptotically analyzable due to the stationary result Theorem~\ref{main-s}.
%The analysis of its NG version can be dissected into two main directions. In the action-to-environment direction, the evolution of price distributions under a given pricing policy can be mapped out through an iterative procedure. In the environment-to-action direction, a firm's optimal pricing policy under a given sequence of price distributions can be solved using ordinary dynamic programming. Under mild assumptions on the demand-arrival behavior, we can use a continuity-based fixed point theorem on the composite action-to-action operator to establish and characterize an equilibrium pricing policy for all firms.

%The application of Proposition~\ref{main} would then result in Theorem~\ref{application}, demonstrating indeed that the time- and inventory-dependent but observation-blind NG-equilibrium pricing policy can be used by all firms in a many-firm situation to generate total payoffs that are hard to beat on average. In contrast, the actual finite-firm situation does not seem to yield easily to analysis. For one thing, as individual players have discernible and random impacts on the evolution of state distributions, no clear-cut definition of the action-to-environment and environment-to-action directions seems possible.

As our foremost contribution, we established one more link between NGs and their finite-game counterparts. Previously, links were mostly established for single-period games, special multi-period games without individual states, or games exhibiting stationary features. The introduction of information-carrying individual states allow in for proper treatment a much wider body of applicable situations involving present-future tradeoffs and transient properties.
%By showing that NG equilibria can be used to generate asymptotic equilibria for ever growing finite games, our message is essentially lower hemi-continuity in nature. It provides one more justification to the study of NGs.
%Second, we show that the NG concept be useful for dynamic games with exogenous global shocks.
Comparing to the earlier work Yang \cite{Y15} which dealt with the NG-finite link for Markovian games as well, the current paper treats more general, non-discrete state and action spaces. As a tradeoff, we are compelled to let random shocks drive both decision making and state evolution. This, as is  backed up by results such as Aumann \cite{A64b} on the interchangeability between the presentations with and without drivers, does not much restrict the generality of our results. Moreover, we demonstrated that the usefulness of SEs to large finite games stems from more fundamental properties possessed by transient NG equilibria. 

Here is our plan for the remainder of the paper. We spend Section~\ref{literature} on a survey of related research and Section~\ref{setup} on basic model primitives. The nonatomic game is  introduced in Section~\ref{games}, while finite games are treated in Section~\ref{finites}. We present the main transient convergence results in Section~\ref{en-converge}. %, and that of equilibria in Section~\ref{eq-converge}.
These are used in Section~\ref{stationary} to establish a link between SEs %in infinite games and equilibria in
and large finite games with stationary features. %We apply our results to dynamic price competition among heterogeneous firms in Section~\ref{example}. 
Further discussion is made in Section~\ref{discussion}, while the paper is concluded in Section~\ref{conclusion}.
%We work on an example in Section~\ref{example}, and conclude the paper in Section~\ref{conclusion}.

\section{Literature Survey}\label{literature}

NGs are often easier to analyze than their finite counterparts, because in them, the action of an individual player has no impact on payoffs and future state evolutions of the other players. Therefore, they are often used as proxies of real competitive systems in economic studies; see, e.g., Aumann \cite{A64a} and Reny and Perry \cite{RP06}.
Systematic research on NG started with Schmeidler \cite{S73}. He formulated a single-period semi-anonymous NG, wherein the joint distribution of other players' types and actions may affect any given player's payoff. When the action space is finite, Schmeidler established the existence of pure equilibria when the game becomes anonymous, so that only the distribution of other players' actions matters.
%A finite-player counterpart to Schmeidler was obtained by Rashid \cite{R83}.
Mas-Colell \cite{M84} showed the existence of distributional equilibria in anonymous NGs with compact action spaces. Khan, Rath, and Sun \cite{KRS99} identified a certain limit to which Schmeidler's result can be extended. Links between NGs and their finite counterparts were covered in Green \cite{G84}, Housman \cite{H88}, Carmona \cite{C04}, Kalai \cite{K04}, Al-Najjar \cite{A08}, and Yang \cite{Y11}.

This paper differs from the above by its focus on multi-period games. For such games without individual states that allow past actions to impact future gains, Green \cite{G80}, Sabourian \cite{S90}, and Al-Najjar and Smorodinsky \cite{AS01} showed that equilibria for large games are nearly myopic.
%Like these works, we deal with large multi-period situations.
With individual states that inherit traces of past actions, the games we study pose new challenges. %Our main message is about observation-blindness.
An NG equilibrium for our situation is certainly not myopic as it takes into account the current action's future consequences. Rather, it is insensitive to real-time observations made on other players' states. We succeed in showing that such a simple action plan can be used profitably in finite situations with randomly evolving state distributions of which a player may have zero, partial, or full knowledge. The type of NGs we deal with are similar to sequential anonymous games studied by Jovanovic and Rosenthal \cite{JR88}, who established existence of distributional equilibria. The result was generalized to games involving aggregate shocks by Bergin and Bernhardt \cite{BB95}.
%As may be seen from Mertens and Parthasarathy \cite{MP88} and Duffie et al. \cite{DGMM94}, equilbrium existence results for finite-player counterparts would require stronger assumptions.
Different from these papers, we work on the link between NGs and finite games, not the NGs themselves.
%; and (iii) the last two papers are not applicable to the searching of pure action plans for the dyanmice pricing game.

In their effort to simplify dynamic games, some authors went further than silencing individual players' influences as done through the NG approach. In addition, they pursued the so-called stationary equilibria (SE), which stressed further the long-run steady-state nature of individual action plans and system-wide state distributions; see, e.g., Hopenhayn \cite{H92} and Adlakha and Johari \cite{AJ10}.
The oblivious equilibrium (OE) concept as proposed by Weintraub, Benkard, and van Roy \cite{WBvR08}, though accounting for impacts of large players, took the same stationary approach by letting firms beware of only long-run average state distributions. We caution that the implicit stationarity of SE or OE renders it inappropriate for applications that are transient by nature; for instance, the dynamic pricing game studied in Yang \cite{Y17} where the inventory level of every firm can only decrease over time.

Some recent works also contributed on the links between equilibria of infinite-player games and their finite-player brethren. Weintraub, Benkard, and van Roy \cite{WBvR11} did so for a setting where long-run average system state can be defined. Adlakha, Johari, and Weintraub \cite{AJW11} established the existence of SE and achieved a similar conclusion by using only exogenous conditions on model primitives.
Weintraub et al. \cite{WBJvR08} studied nonstationary oblivious equilibria (NOE) that capture transient behaviors of players, and showed their usefulness in finite-player situations by relying on a ``light-tail'' condition on players' state distributions similar to that used in \cite{WBvR11}. Huang, Malhame, and Caines \cite{HMC06} dealt with a continuous-time multi-player system where independent diffusion processes provide random drivers. They reached equilibria in the nonatomic limit, and derived asymptotic results as the number of players becomes large. In the work, other players impact a given player through linear functionals of the state distribution they form; meanwhile, their actions play no direct role. 

Our discrete-time framework afforded us almost full generality regarding other players' impacts on any given player's payoffs and state transitions---it is the joint state-action distribution that forms the environment faced by an individual player.  %Based on a model involving aggregate states and discounting, %and mixed strategies,
%Bodoh-Creed \cite{BC12} reached results similar in spirit to our Theorem~\ref{main} (his Theorem 2). %Certain assumptions like the one on the bounded radius of individial-state evolution (his Assumption 2) are not required by us. Also, h
%His notion of semi-anonymity omits the copula that links the two marginal distributions on states and actions, respectively. %Also, the equilibrium action profiles he considered depend on individual-state distributions. The latter we think goes against the purpose of simplifying matters by considering NG equilibria.
As already mentioned, while this paper tackles the case where exogenous shocks drive state evolution and decision making, Yang \cite{Y15} dealt with the setting where such shocks are not necessarily identifiable; however, technical challenges faced there forced state and action spaces to be discrete.

\section{Model Primitives}\label{setup}

In the dynamic games we study, players are engaged in multi-period competition in periods $t=1,2,...,\bar t$. In period $t$, a player's payoff $\psi_t(s,x,\mu)$ depends on his state $s$, action $x$, and some $\mu$ depicting the outside environment. We suppose the game is semi-anonymous, so that $\mu$ can just be the joint distribution of other players' states and actions. %Note that the state $s$ can contain a player's type $s_T$, such as a firm's efficiency level, which does not shift from one period to another, as well as his moving status $s_I$, such as a firm's inventory level, which does. 
The dynamics of the game is represented by a function $\theta_t(s,x,\mu,i)$, where $s$, $x$, and $\mu$ are defined as above, and $i$ is an idiosyncratic shock the player experiences individually after taking his action. All players' post-action shocks are independently sampled from a common distribution $\iota$.

We allow players to cast dices to decide their actions. However, we do not model the extent to which players can observe their outside environments; after all, we focus only on action plans that do not take advantage of any such observations. In every period $t$, we suppose each player receives another idiosyncratic shock $g$, this time before taking his action. All players' pre-action shocks, such as outcomes of dice casts, are independently sampled from a common distribution $\gamma$.
%These shocks act like outcomes of dices the players throw to determine their actions.
We study the case where a player's action $x_t(s,g)$ depends merely on his own state $s$ and the shock $g$ that he himself has received. The main purpose of the paper is to show that one such action plan $x_{[1\bar t]}\equiv (x_t)_{t=1,...,\bar t}$ is quite sufficient for the multi-period game just described, even when the latter may be transient in nature and of the more complex finite-player variety.

%Our payoff and state transition functions $\psi_t$ and $\theta_t$ allow general dependences on the joint distribution $\mu$ of other players' states and actions. This makes our conclusions applicable to the competitive dynamic price situation studied in Yang \cite{Y17}. We will point out in Section~\ref{discussion} where our model is more general than existing ones, such as those studied by Weintraub et al. \cite{WBJvR08}, Weintraub, Benkard, and van Roy \cite{WBvR11}, and Huang, Malhame, and Caines \cite{HMC06}. 

Some notations are needed for formal definitions.  
Given a metric space $A$, we use $d_A$ to denote its metric, ${\cal B}(A)$ its Borel $\sigma$-field, and ${\cal P}(A)$ the set of all probability measures (distributions) on the measurable space $(A,{\cal B}(A))$. The space ${\cal P}(A)$ is metrized by the Prohorov metric $\rho_A$, which induces on it the weak topology.
%and the resultant metric space is separable too (Ethier and Kurtz \cite{EK86}, Theorem III.1.7).
%The aforementioned joint state-action distribution comes from the space ${\cal P}(S\times X)$.
Given metric spaces $A$ and $B$, we use ${\cal M}(A,B)$ to represent all measurable functions from $A$ to $B$. 

We use complete separable metric space $S$ for individual states $s$ and separable metric space $X$ for player actions $x$.
%Every time players are engaged, each is aware of his own individual state $s\in S$.
%When their number becomes large, players' knowledge on others' states will become ever less revelant.
In a semi-anonymous fashion, payoffs and state transitions depend on the joint distribution $\mu\in {\cal P}(S\times X)$ of other players' states and actions.
%We let shocks drive players' decision making processes as well as their random state transitions. That is, p
Let pre-action shocks $g$ come from a complete separable metric space $G$. In every period, these action-influencing shocks are independently drawn from a common distribution $\gamma\in {\cal P}(G)$. Let post-action shocks $i$ come from a complete separable metric space $I$. In every period, these transition-influencing shocks are independently drawn from a common distribution $\iota\in {\cal P}(I)$. The completeness requirements on $S$, $G$, and $I$ stem from the need to invoke Lemma~\ref{joint} in Appendix~\ref{app-a-en}. These are certainly not stringent. %, as multi-dimensional Euclidean spaces, as well as finite and countable sets endowed with the discrete metric,
%where every two different points has a distance 1, are all complete separable metric spaces.

For period $t=1,...,\bar t$, a player's state $s\in S$, his action $x\in X$, and the joint state-action distribution $\mu\in {\cal P}(S\times X)$ he faces, together determine his payoff in period $t$. In fact, we require there to be a bounded payoff function
\begin{equation}\label{d-payoff}
\psi_t:S\times X\times {\cal P}(S\times X)\longrightarrow [-\bar\psi_t,\bar\psi_t],
\end{equation}
where $\bar\psi_t$ is some positive constant. It satisfies that $\psi_t(\cdot,\cdot,\mu)\in {\cal M}(S\times X,[-\bar\psi_t,\bar\psi_t])$ for every given distribution $\mu\in {\cal P}(S\times X)$. %Note that the function may itself be an average over different shocks $i$.
As the same player will enter a new state under post-action shock $i\in I$, we require there to be 
\begin{equation}\label{d-transition}
\theta_t:S\times X\times {\cal P}(S\times X)\times I\longrightarrow S.
\end{equation}
It satisfies that $\theta_t(\cdot,\cdot,\mu,\cdot)\in {\cal M}(S\times X\times I,S)$ at every distribution $\mu\in {\cal P}(X\times S)$.

The action plans we consider are of the form
\begin{equation}\label{d-action}
x_t:S\times G\longrightarrow X,
\end{equation}
which are required to be members of ${\cal M}(S\times G,X)$. That is, a player will use action $x_t(s,g)$ in period $t$ when he starts with state $s\in S$ and receives pre-action shock $g$. %We can always use a singleton pre-action shock space $G$ and hence degenerate $\gamma$ to handle pure action plans and use a singleton post-action shock space $I$ and hence degenerate $\iota$ to handle deterministic state transitions. 
We call any state distribution $\sigma\in {\cal P}(S)$ a pre-action environment, because the one formed by other players is what a player could potentially see at the beginning of any period. Also, call any joint state-action distribution $\mu\in {\cal P}(S\times X)$ an in-action environment, because the one formed by other players is what a player could potentially see in the midst of play in any period.

Let us recount our model primitives as follows: the horizon length $\bar t$; the state space $S$, the action space $X$, the pre-action shock space $G$, the post-action shock space $I$; also, the pre-action shock distribution $\gamma$, the post-action shock distribution $\iota$; finally for periods $t=1,...,\bar t$, the payoff functions $\psi_t$ and state transition functions $\theta_t$. 
 
\section{The Nonatomic Game}\label{games}

Given an initial pre-action environment $\sigma_1\in {\cal P}(S)$, we can define a nonatomic game $\Gamma(\sigma_1)$ which starts period 1 with $\sigma_1$ as the distribution of all players' states. We focus on policy profiles of the form $x_{[1\bar t]}\equiv (x_t)_{t=1,...,\bar t}\in ({\cal M}(S\times G,X))^{\bar t}$, where each $x_t\in {\cal M}(S\times G,X)$ is a map from a player's state-shock pairs to actions. Along with the given initial environment $\sigma_1$, we suppose such a profile will help generate a deterministic pre-action environment trajectory $\sigma_{[1,\bar t+1]}\equiv(\sigma_t)_{t=1,2,...,\bar t,\bar t+1}\in ({\cal P}(S))^{\bar t+1}$. This allows a player's policy to be observation-blind; that is, what portion of $\sigma_t$ is observable to the player in each period $t$ is not of any concern. The determinism of the environment evolution in $\Gamma(\sigma_1)$ is justifiable by Sun's \cite{S06} LLN involving a continuum of indexed players. 
% evolution in $\Gamma(\sigma_1)$ can be justified by an exact LLN involving a heuristically argued along the line of LLN. However, in view of Feldman and Giles \cite{FG85} and Judd \cite{J85}, a rigorous proof does not seem to be in the offing. {\bf Sun \cite{S06} blah blah.  Thus, we have made this feature an axiom of the nonatomic game rather than a property derivable from its finite counterparts.} %We believe this adds to the weight of our main result instead of the opposite. It can now be said that, a solution to an NG, model-wise not necessarily representing the limiting behavior of a continuum of players in the most rigorous sense, can nevertheless be shown through LLN to offer proper solutions to large finite games.

We now discuss how the deterministic trajectory can be formed. Let $t=1,...,\bar t$ be given.
%For a player with starting state $s_t$, his action will be $x_t(s_t,g_t)$ after receiving his pre-action shock $g_t$.
When all players form state distribution $\sigma_t\in {\cal P}(S)$ at the beginning and adopt the same plan $x_t\in {\cal M}(S\times G,X)$ for the period, the in-action environment $\mu_t\equiv M(\sigma_t,x_t)\in {\cal P}(S\times X)$ to be experienced by all players will take the form
\begin{equation}\label{s-to-mu}
\mu_t=M(\sigma_t,x_t)=(\sigma_t\times\gamma)\cdot (\mbox{prj}_S,x_t)^{-1},
\end{equation}
where $\mbox{prj}_S$ stands for the projection map from $S\times  G$ to $S$. The meaning for~(\ref{s-to-mu}) is that, for any measurable joint state-action set $W'\in {\cal B}(S\times X)$,
\begin{equation}\label{yanyi}
\mu_t(W')=(\sigma_t\times\gamma)\left((\mbox{prj}_S,x_t)^{-1}(W')\right)=\int_S\int_G{\bf 1}[(s,x_t(s,g))\in W']\cdot\gamma(dg)\cdot\sigma_t(ds).
%\int_G\sigma_t(\{s\in S\mid (s,x_t(s,g))\in W'\})\cdot\gamma(dg).
\end{equation}
This reflects that the joint distribution for states and pre-action shocks is the product form  $\sigma_t\times\gamma$; also, $x_t$ provides the map from state-shock pairs to actions for this period.

For a player who starts with state $s_t$ and has experienced pre-action shock $g_t$ as well as post-action shock $i_t$, his new state will be governed by~(\ref{d-transition}):
\begin{equation}\label{sdf}
s_{t+1}=\theta_t\left(s_t,x_t(s_t,g_t),M(\sigma_t,x_t),i_t\right).
\end{equation}
%Here, for $a=(a_1,...,a_n)\in A^n$ and $m=1,2,...,n$, we have used $a_{-m}$ to represent $(a_1,...,a_{m-1},a_{m+1},...,a_n)$.
To describe the transition of the overall pre-action environment from $\sigma_t$ to $\sigma_{t+1}$ under action plan $x_t$, we define operator $T_t(x_t)$ on ${\cal P}(S)$. Note that  states are distributed according to $\sigma_t$, pre-action shocks are distributed according to $\gamma$, and
post-action shocks are distributed according to $\iota$. So following~(\ref{sdf}),
\begin{equation}\label{T-def}
\sigma_{t+1}=T_t(x_t)\circ \sigma_t=(\sigma_t\times\iota\times\gamma)\cdot \left[\theta_t\left(\mbox{prj}_S,x_t\cdot \mbox{prj}_{S\times G},M(\sigma_t,x_t),\mbox{prj}_I\right)\right]^{-1},
\end{equation}
meaning that, for any measurable action set $S'\in {\cal B}(S)$,
\begin{equation}\label{T-def0}\begin{array}{l}
\sigma_{t+1}(S')=[T_t(x_t)\circ \sigma_t](S')\\
\;\;\;\;\;\;\;\;\;\;\;\;=\int_S\int_{G}\int_I {\bf 1}\left[\theta_t(s,x_t(s,g),M(\sigma_t,x_t),i)\in S'\right]\cdot\iota(di)\cdot \gamma(dg)\cdot\sigma_t(ds).
\end{array}\end{equation}
We can iteratively define $T_{[tt']}(x_{[tt']})$ for $t'=t-1,t,t+1,...$ so that $T_{[t,t-1]}$ is the identity mapping on ${\cal P}(S)$ and for $t'=t,t+1,...$,
\begin{equation}\label{it-0}
T_{[tt']}(x_{[tt']})=T_{t'}(x_{t'})\circ T_{[t,t'-1]}(x_{[t,t'-1]}).
\end{equation}
The environment trajectory alluded to earlier is therefore
\begin{equation}\label{sequence}
\sigma_{[1,\bar t+1]}=(T_{[1,t-1]}(x_{[1,t-1]})\circ\sigma_1)_{t=1,2,...,\bar t,\bar t+1}.
\end{equation}
%It is deterministic by design.

%\section{NG Equilibria}\label{news}

%Note the absence of $x_{\bar t}$ in the creation of the sequence $\sigma$. This last action plan, along with $\sigma_{\bar t}$, will determine the terminal pre-action environment $\sigma_{\bar t+1}$. But the latter is inconsequential due to the zero terminal-value assumption.

In defining $\Gamma(\sigma_1)$'s equilibria, we subject a candidate policy profile to the one-time deviation of a single player, who is negligible in his influence over others.
%, rather than one-time deviations made by players of the same state who may make up a substantial body.
%our definition has nothing to do with whether or not the concerned distributions $\sigma_t$ are nonatomic; also,
The deviation will not alter the environment trajectory corresponding to the candidate profile. Thus, we define $v_t(s_t,\sigma_t,x_{[t\bar t]},y_t)$ as the total expected payoff a player can make from time $t$ to $\bar t$, when he starts with state $s_t\in S$, other players form pre-action environment $\sigma_t\in {\cal P}(S)$, all players adopt policy $x_{[t\bar t]}\equiv(x_{t'})_{t'=t,...,\bar t}\in ({\cal M}(S\times G,X))^{\bar t-t+1}$ with the exception of the current player in period $t$ alone, who deviates to policy $y_t\in {\cal M}(S\times G,X)$ in that period. As a terminal condition, we have
\begin{equation}\label{terminal}
v_{\bar t+1}(s_{\bar t+1},\sigma_{\bar t+1},y_{\bar t+1})=0.
\end{equation}
For $t=\bar t,\bar t-1,...,1$, we have
\begin{equation}\label{recursive}\begin{array}{ll}
v_t(s_t,\sigma_t,x_{[t\bar t]},y_t)=\int_G[\psi_t(s_t,y_t(s_t,g_t),M(\sigma_t,x_t))+\int_I v_{t+1}(\theta_t(s_t,y_t(s_t,g_t),\\
\;\;\;\;\;\;\;\;\;\;\;\;\;\;\;\;\;\;\;\;\;\;\;\;\;\;\;\;\;\;\;\;\;\;\;\;M(\sigma_t,x_t),i_t),T_t(x_t)\circ\sigma_t,x_{[t+1,\bar t]},x_{t+1})\cdot\iota(di_t)]\cdot\gamma(dg_t),
	\end{array}\end{equation}
due to the dynamics illustrated in~(\ref{sdf}) to~(\ref{T-def0}). The deviation $y_t$ affects the current player's action $y_t(s_t,g_t)$ in period $t$ and his own state $\theta_t(s_t,y_t(s_t,g_t),M(\sigma_t,x_t),i_t)$ in period $t+1$. But as a distinctive feature of the NG setup, it has no bearing on the period-$(t+1)$ pre-action environment $T_t(x_t)\circ\sigma_t$.
	
Now define $u_t:{\cal P}(S)\times ({\cal M}(S\times G,X))^{\bar t-t+1}\times {\cal M}(S\times G,X)\longrightarrow \Re$ so that
\begin{equation}\label{ut-def}
u_t(\sigma_t,x_{[t\bar t]},y_t)=\int_S v_t(s_t,\sigma_t,x_{[t\bar t]},y_t)\cdot \sigma_t(ds_t).
\end{equation}
This can be understood as one particular player's average gain from period $t$ onward when the same conditions specified earlier prevail and his period-$t$ state is sampled from the distribution $\sigma_t$.
We deem policy $x^*_{[1\bar t]}\equiv (x^*_t)_{t=1,2,...,\bar t}\in ({\cal M}(S\times G,X))^{\bar t}$ a Markov equilibrium for the game $\Gamma(\sigma_1)$ when, for every $t=1,2,...,\bar t$ and $y_t\in {\cal M}(S\times G,X)$,
\begin{equation}\label{first}
u_t\left(T_{[1,t-1]}(x^*_{[1,t-1]})\circ \sigma_1,x^*_{[t\bar t]},x^*_t\right)\geq u_t\left(T_{[1,t-1]}(x^*_{[1,t-1]})\circ \sigma_1,x^*_{[t\bar t]},y_t\right).
\end{equation}
That is, policy $x^*_{[1\bar t]}$ will be regarded an equilibrium when it cannot be bettered by any plan $y_t\in {\cal M}(S\times G,X)$ in any period $t$ in an average sense that is defined by the period-$t$ environment $\sigma_t=T_{[1,t-1]}(x^*_{[1,t-1]})\circ \sigma_1$. We caution that~(\ref{first}) is weaker than
\begin{equation}\label{weaker}\begin{array}{l}
v_t\left(s_t,T_{[1,t-1]}(x^*_{[1,t-1]})\circ \sigma_1,x^*_{[t\bar t]},x^*_t\right)\geq v_t\left(s_t,T_{[1,t-1]}(x^*_{[1,t-1]})\circ \sigma_1,x^*_{[t\bar t]},y_t\right),
\end{array}\end{equation}
for every $s_t\in S$. On the other hand, since $y_t\in {\cal M}(S\times G,X)$ allows for much freedom in choosing for each state $s\in S$ and shock $g\in G$ a competitive reaction $y_t(s,g)$, there is not much difference between the two criteria aside from measurability subtleties.

\section{Finite-player Games}\label{finites}

More notations are needed to appropriately describe finite games. For metric space $A$ and $a\in A$, we use $\varepsilon(a)$ to denote the singleton probability measure with $\varepsilon(a)(\{a\})=1$. For $a=(a_1,...,a_n)\in A^n$ where $n\in\mathbb{N}$, the set of natural numbers, we use $\varepsilon(a)$ for $\sum_{m=1}^n \varepsilon(a_m)/n$. The two uses are consistent. We also use ${\cal P}_n(A)$ for the space of probability measures of the type $\varepsilon(a)$ for $a\in A^n$, i.e., the space of empirical distributions generated from $n$ samples.

For some $n=2,3,...$ and initial state distribution $\sigma_1\in {\cal P}_n(S)$, we can define an $n$-player game $\Gamma_n(\sigma_1)$. Note the initial pre-action environment $\sigma_1$ must be of the form $\varepsilon(s_1)=\varepsilon(s_{11},s_{12},...,s_{1n})$, where each $s_{1m}\in S$ is player $m$'s initial state. The game's payoffs and state transitions are still governed by~(\ref{d-payoff}) and~(\ref{d-transition}), respectively. In period $t$, the pre-action environment is also some $\sigma_t=\varepsilon(s_{t1},...,s_{tn})\in {\cal P}_n(S)\subset {\cal P}(S)$. Hence, the in-action environment $\mu_{t1}\in {\cal P}_{n-1}(S\times X)\subset {\cal P}(S\times X)$ experienced by any designated player 1 is the empirical distribution $\varepsilon(s_{t,-1},y_{t,-1})=\varepsilon((s_{t2},y_{t2}),...,(s_{tn},y_{tn}))$ when each player $m$ is in state $s_{tm}\in S$ and takes action $y_{tm}\in X$. Let players still adopt policy $x_{[1\bar t]}\equiv (x_t)_{t=1,...,\bar t}\in ({\cal M}(S\times G,X))^{\bar t}$, which is but the crudest of many choices available to the $n$ players. We shall see later that this restriction is not going to do much harm.

Simplistic as it may seem, $x$ will not merely generate a deterministic environment trajectory. Given pre-action shock vector $g_t=(g_{t1},...,g_{tn})\in G^n$ and post-action shock vector $i_t=(i_{t1},...,i_{tn})\in I^n$, we can define $T_{nt}(x_t,g_t,i_t)$ as the operator on ${\cal P}_n(S)$ that converts a period-$t$ pre-action environment into a period-$(t+1)$ one. Thus following~(\ref{s-to-mu}) to~(\ref{sdf}), $\varepsilon(s_{t+1})=T_{nt}(x_t,g_t,i_t)\circ \varepsilon(s_t)$ is such that
\begin{equation}\label{Tn-1-def}
s_{t+1,m}=\theta_t\left(s_{tm},x_t(s_{tm},g_{tm}),M_n(s_{t,-m},g_{t,-m},x_t),i_{tm}\right),\hspace*{.5in}\forall m=1,2,...,n,
\end{equation}
where
\begin{equation}\label{s-to-mu-new}
M_n(s_{t,-m},g_{t,-m},x_t)=\varepsilon(s_{t,-m},g_{t,-m})\cdot (\mbox{prj}_S,x_t)^{-1},
\end{equation}
and each $\varepsilon(s_{t,-m},g_{t,-m})$ represents the empirical distribution built on state-shock pairs $(s_{t1},g_{t1})$, $...$, $(s_{t,m-1},g_{t,m-1})$, $(s_{t,m+1},g_{t,m+1})$, $...$, $(s_{tn},g_{tn})$.
%; consequently, each $M_n(s_{t,-m},g_{t,-m},x_t)$ represents the empirical distribution built on state-action pairs $(s_{t1},x_t(s_{t1},g_{t1}))$, $...$, $(s_{t,m-1},x_t(s_{t,m-1},g_{t,m-1}))$, $(s_{t,m+1},x_t(s_{t,m+1},g_{t,m+1}))$, $...$, $(s_{tn},x_t(s_{tn},g_{tn}))$. 
The latter reflects that player $m$'s in-action environment is made up of the states and actions of the other $n-1$ players along with the common action plan adopted by all players. Again, we define $T_{n,[tt']}$ as the identity map when $t'\leq t-1$ and when $t\leq t'$, let
\begin{equation}\label{it-n}
T_{n,[tt']}(x_{[tt']},g_{[tt']},i_{[tt']})=T_{nt'}(x_{t'},g_{t'},i_{t'})\circ T_{n,[t,t'-1]}(x_{[t,t'-1]},g_{[t,t'-1]},i_{[t,t'-1]}).
\end{equation}
The evolution of pre-action envirnoments $\sigma_t=\varepsilon(s_t)$ is guided by the random shock vectors $g_t$ and $i_t$, and hence is stochastic by nature.

For an $n$-player game, let $v_{nt}(s_{t1},\varepsilon(s_{t,-1}),x_{[t\bar t]},y_t)$ be the total expected payoff player 1 can make from $t$ to $\bar t$, when he starts with state $s_{t1}\in S$, other players' initial environments are describable by their aggregate empirical state distribution $\varepsilon(s_{t,-1})=\varepsilon(s_{t2},...,s_{tn})$, and all players adopt the policy $x_{[t\bar t]}\equiv(x_{t'})_{t'=t,...,\bar t}\in ({\cal M}(S\times G,X))^{\bar t-t+1}$ from period $t$ to period $\bar t$ with the exception of player 1 in period $t$ alone, who deviates to policy $y_t\in {\cal M}(S\times G,X)$. As a terminal condition, we have
\begin{equation}\label{terminal-n}
v_{n,\bar t+1}(s_{\bar t+1,1},\varepsilon(s_{\bar t+1,-1}),y_{\bar t+1})=0.
\end{equation}
For $t=\bar t,\bar t-1,...,1$, we have the recursive relationship
\begin{equation}\label{recursive-n}\begin{array}{l}
v_{nt}(s_{t1},\varepsilon(s_{t,-1}),x_{[t\bar t]},y_t)=\int_{G^n}\gamma^n(dg_t)\times\{\psi_t(s_{t1},y_t(s_{t1},g_{t1}),M_n(s_{t,-1},g_{t,-1},x_t))\\
\;\;\;\;\;\;\;\;\;\;\;\;\;\;\;\;\;\;+\int_{I^n}\iota^n(di_t)\times v_{n,t+1}(\theta_t(s_{t1},y_t(s_{t1},g_{t1}),M_n(s_{t,-1},g_{t,-1},x_t),i_{t1}),\\
\;\;\;\;\;\;\;\;\;\;\;\;\;\;\;\;\;\;\;\;\;\;\;\;\;\;\;\;\;\;\;\;\;\;\;\;\;\;\;\;\;\;\;\;\;\;\;\;[T_{nt}(x_t,g_t,i_t)\circ \varepsilon(s_t)]_{-1},x_{[t+1,\bar t]},x_{t+1})\},
\end{array}\end{equation}
due to the dynamics illustrated in~(\ref{sdf}) and~(\ref{Tn-1-def}). By $[T_{nt}(x_t,g_t,i_t)\circ \varepsilon(s_t)]_{-1}$, we mean $\varepsilon(s_{t+1,-1})$, where $\varepsilon(s_{t+1})$ is $T_{nt}(x_t,g_t,i_t)\circ \varepsilon(s_t)$ as defined through~(\ref{Tn-1-def}). The current~(\ref{recursive-n}) is much more complicated than the NG counterpart~(\ref{recursive}). The evolution from period $t$ to $t+1$ now depends on pre-action  shocks $g_t\equiv (g_{t1},...,g_{tn})$ and post-action shocks $i_t\equiv (i_{t1},...,i_{tn})$. Also, the in-action environment $M_n(s_{t,-1},g_{t,-1},x_t)$ experienced by player 1 excludes his own state and action, and hence is different from the environment faced by any other player. Similarly, the in-action environment $[T_{nt}(x_t,g_t,i_t)\circ \varepsilon(s_t)]_{-1}$ to be faced by player 1 in period $t+1$ is unique to him as well. The added complexity motivates us to exploit the easier-to-handle NG case. %This entity depends on $i_t$ only through $i_{t,-1}=(i_{t2},...,i_{tn})$ and has nothing to do with whether action rule $x_t$ has been taken with player 1's state $s_{t1}$.

Let $\sigma_{[1\bar t]}\equiv(\sigma_t)_{t=1,...,\bar t}\in ({\cal P}(S))^{\bar t}$ be a sequence of environments. For $\epsilon\geq 0$, we deem $x^*_{[1\bar t]}\equiv(x^*_t)_{t=1,...,\bar t}\in ({\cal M}(S\times G,X))^{\bar t}$ an $\epsilon$-Markov equilibrium for the game family $(\Gamma_n(\varepsilon(s_1))\mid s_1\in S^n)$ in the sense of $\sigma_{[1\bar t]}$ when, for every $t=1,...,\bar t$ and $y_t\in {\cal M}(S\times G,X)$,
\begin{equation}\label{second}
\int_{S^n}v_{nt}\left(s_{t1},\varepsilon(s_{t,-1}),x^*_{[t\bar t]},x^*_t\right)\cdot \sigma^{\;n}_t(ds_t)\geq \int_{S^n}v_{nt}\left(s_{t1},\varepsilon(s_{t,-1}),x^*_{[t\bar t]},y_t\right)\cdot \sigma^{\;n}_t(ds_t)-\epsilon.
\end{equation}
That is, action plan $x^*_{[1\bar t]}$ will be an $\epsilon$-Markov equilibrium in the sense of $\sigma_{[1\bar t]}$ when under its guidance, the average payoff from any period $t$ on will not be improved by more than $\epsilon$ through any deviation, where the ``average'' is taken with respect to state distribution $\sigma_t$.

\section{Main Convergence Results}\label{en-converge}

We can achieve convergences of environments and then of equilibria. The former is more fundamental and challenging, and the latter is built on it.

\subsection{Convergence of Environments}\label{obama}

Even without touching upon payoffs or equilibria, we can establish a link between finite games and their NG counterpart. It reflects that stochastic environment pathways experienced by large finite games converge to the NG's deterministic environment trajectory.
%For this, we may denote by $\tilde S_t(s,x,\mu)$ the random state defined on the probability space $(I,{\cal B}(I),\iota)$ that maps each $i\in I$ to $\theta_t(s,x,\mu,i)\in S$.

Let $A$, $B$, and $C$ be metric spaces and $\pi_B\in {\cal P}(B)$ be a distribution. We use ${\cal K}(A,B,\pi_B,C)$ $\subseteq {\cal M}(A\times B,C)$ to represent the space of all measurable functions from $A\times B$ to $C$ that are uniformly continuous in a probabilistic sense. The criterion for $y\in {\cal K}(A,B,\pi_B,C)$ is that for any $\epsilon>0$, there exists $\delta>0$, so that for any $a,a'\in A$ satisfying $d_A(a,a')<\delta$,
\begin{equation}\label{wawawa}
\pi_B\left(\{b\in B|d_C(y(a,b),y(a',b))<\epsilon\}\right)>1-\epsilon.
\end{equation}
When $B$ is a singleton and hence $\pi_B$ is degenerate, $y\in {\cal K}(A,B,\pi_B,C)$ merely means that $y$ is a uniformly continuous function from $A$ to $C$, a situation we denote by $y\in {\cal K}(A,C)$. For regular $B$ and $\pi_B$, the meaning is somehow that continuity will happen in most cases. 
%Probabilistic continuity has another much anticipated link with ordinary continuity; see Lemma~\ref{curiosity} of Appendix~\ref{app-a-en}. It shows that, as a requirement, $y\in {\cal K}(A,B,\pi_B,C)$ is stronger than $\upsilon(a)=\pi_B\cdot (y(a,\cdot))^{-1}$ being uniformly continuous in $a$, but probably not by much.

We now make two assumptions on the transition function $\theta_t$:\\
\indent\M (S1) For every $\mu\in {\cal P}(S\times X)$, the function $\theta_t(\cdot,\cdot,\mu,\cdot)$ is a member of ${\cal K}(S\times X,I,\iota,S)$. That is, for any $\mu\in {\cal P}(S\times X)$ and $\epsilon>0$, there exist $\delta_S>0$ and $\delta_X>0$, so that for any $s,s'\in S$ and $x,x'\in X$ satisfying $d_S(s,s')<\delta_S$ and $d_X(x,x')<\delta_X$,
\[ \iota\left(\{i\in I\mid d_S(\theta_t(s,x,\mu,i),\theta_t(s',x',\mu,i))<\epsilon\}\right)>1-\epsilon.\]
\indent\M (S2) Not only is it true that $\theta_t(s,x,\cdot,\cdot)\in {\cal K}({\cal P}(S\times X),I,\iota,S)$ at every $(s,x)\in S\times X$, but the continuity is also achieved at a rate independent of the $(s,x)$ present. That is, for any $\mu\in {\cal P}(S\times X)$ and $\epsilon>0$, there exists $\delta>0$, so that for any $\mu'\in {\cal P}(S\times X)$ satisfying $\rho_{S\times X}(\mu,\mu')<\delta$, as well as any $s\in S$ and $x\in X$,
\[\iota(\{i\in I\mid d_S(\theta_t(s,x,\mu,i),\theta_t(s,x,\mu',i))<\epsilon\})>1-\epsilon.\]

%From the detailed study of the dynamic pricing game in Section~\ref{example}, we shall see that these requirements are easy to meet. Now f
For separable metric space $A$, we use $(A^n,{\cal B}^n(A))$ to denote the product measurable space that houses $n$-long sample sequences. Given $\pi\in {\cal P}(A)$, we use $\pi^n$ to denote the product measure on $(A^n,{\cal B}^n(A))$. We can show that a one-step evolution in a big game is not that much different from that in a nonatomic game.

\begin{proposition}\label{T-onestep}
	Given separable metric space $A$, distribution $\pi\in {\cal P}(A)$, and pre-action environment $\sigma\in {\cal P}(S)$, suppose $s_n=(s_n(a)\mid a\in A^n)$ for each $n\in \mathbb{N}$ is a member of ${\cal M}(A^n,S^n)$, and $\varepsilon(s_n(a))$ converges to $\sigma$ in probability, to the effect that
	\[ \pi^n(\{a\in A^n\mid \rho_S(\varepsilon(s_n(a)),\sigma)<\epsilon\})>1-\epsilon,\]
	for any $\epsilon>0$ and any $n$ large enough. Then, any $T_{nt}(x,g,i)\circ \varepsilon(s_n(a))$ will converge to $T_t(x)\circ \sigma$ in probability for any probabilistically continuous $x$. That is, for any $x\in {\cal K}(S,G,\gamma,X)$,
	\[ (\pi\times\gamma\times\iota)^n\left(\{(a,g,i)\in (A\times G\times I)^n\mid \rho_S(T_{nt}(x,g,i)\circ \varepsilon(s_n(a)),T_t(x)\circ \sigma)<\epsilon\}\right)>1-\epsilon, \]
	for any $\epsilon>0$ and any $n$ large enough.
\end{proposition}

Recall that $\rho_S$ is the Prohorov metric for measuring the distance between two state distributions. Also, the operator $T_t(x)$ delineating the period-$t$ transition of an NG's pre-action environment is defined at~(\ref{T-def0}), and its finite-game counterpart $T_{nt}(x,g,i)$ is defined at~(\ref{Tn-1-def}).   The proof of Proposition~\ref{T-onestep}  calls upon Lemma~\ref{joint} in Appendix~\ref{app-a-en}. This is why the spaces $S$, $G$, and $I$ are required to be complete. Now imagine that $(A,{\cal B}(A),\pi)$ provides exogenous shocks that drive games' evolutions up to period $t$: $A=S\times G^{t-1}\times I^{t-1}$ and $\pi=\sigma_1\times \gamma^{t-1}\times\iota^{t-1}$. Proposition~\ref{T-onestep} states that, when starting period $t$ with initial state vectors $s_n(a)$ in $n$-player games that in aggregation increasingly resemble the given starting distribution $\sigma$ for the NG, one will still get state vectors in large games that in aggregation resemble the NG's state distribution after the period-$t$ transition. When exploiting this proposition iteratively, we can arrive at our first main result on the convergence of environments.

\begin{theorem}\label{T-converge}
Let a policy profile $x_{[t\bar t]}\in ({\cal M}(S\times G,X))^{\bar t-t+1}$ for periods $t,t+1,...,\bar t$ be such that each $x_{t'}$  is a member of ${\cal K}(S,G,\gamma,X)$.
Then, when we sample $s_t=(s_{t1},...,s_{tn})$ from a given pre-action environment $\sigma_t\in {\cal P}(S)$, the sequence $(\sigma_{nt'})_{t'=t,t+1,...,\bar t,\bar t+1}$ of stochastic pre-action environments will converge to the sequence $(\sigma_{t'})_{t'=t,t+1,...,\bar t,\bar t+1}$ of deterministic pre-action environments in probability, where for each $t'=t,t+1,...,\bar t,\bar t+1$, $\sigma_{nt'}$ is a sample over the $\varepsilon(s_{t'})$'s with $\varepsilon(s_{t'})=T_{n,[t,t'-1]}(x_{[t,t'-1]},g_{[t,t'-1]},i_{[t,t'-1]})\circ \varepsilon(s_t)$, 
while $(s_t,g_{[t,t'-1]},i_{[t,t'-1]})$ is distributed according to $(\sigma_t\times \gamma^{t'-t}\times \iota^{t'-t})^n$; also, $\sigma_{t'}=T_{[t,t'-1]}(x_{[t,t'-1]})\circ\sigma_t$.
 That is, for any $\epsilon>0$ and any $n$ large enough,
	\[ \left(\sigma_t\times\gamma^{\bar t-t+1}\times \iota^{\bar t-t+1}\right)^n\left(\tilde A_n(\epsilon)\right)>1-\epsilon,\]
	where $\tilde A_n(\epsilon)\in {\cal B}^n(S\times G^{\bar t-t+1}\times I^{\bar t-t+1})$ is such that, for any $(s_t,g_{[t,\bar t]},i_{[t,\bar t]})\in \tilde A_n(\epsilon)$,
	\[ \rho_S\left(\sigma_{nt'},\sigma_{t'}\right)<\epsilon,\hspace*{.5in}\forall t'=t,t+1,...,\bar t,\bar t+1. \]
\end{theorem}

The multi-period transition operator $T_{[t,t'-1]}(x_{[t,t'-1]})$ for the NG is defined at~(\ref{it-0}), and its finite-game counterpart $T_{n,[t,t'-1]}(x_{[t,t'-1]},g_{[t,t'-1]},i_{[t,t'-1]})$ is defined at~(\ref{it-n}). Suppose an NG starts period $t$ with pre-action environment $\sigma_t$ and a slew of finite games start the period with pre-action environments that are sampled from $\sigma_t$. Let the evolution of both types of games be guided by players acting according to the same probabilistically continuous policy profile $x_{[t\bar t]}$. Then, as the numbers of players $n$ involved in finite games grow to infinity, Theorem~\ref{T-converge} predicts for ever less chances for the finite games' period-$t'$ environments $\sigma_{nt'}=T_{n,[t,t'-1]}(x_{[t,t'-1]},g_{[t,t'-1]},i_{[t,t'-1]})\circ \varepsilon(s_t)$ to veer off even slightly away from the NG's deterministic period-$t'$ environment $\sigma_{t'}=T_{[t,t'-1]}(x_{[t,t'-1]})\circ\sigma_t$.

\subsection{Convergence of Equilibria}\label{eq-converge}

We now set out to establish this section's main result, that an equilibrium from the NG will serve as an ever more accurate approximate equilibrium for ever larger finite games. First, we need to assume that the single-period payoff functions $\psi_t$ are continuous:\\
\indent\M (F1) Each $\psi_t(s,x,\mu)$ is continuous in $(s,x)$. That is, for any $\mu\in {\cal P}(S\times X)$ and $\epsilon>0$, there exist $\delta_S>0$ and $\delta_X>0$, so that for any $s,s'\in S$ and $x,x'\in X$ satisfying $d_S(s,s')<\delta_S$ and $d_X(x,x')<\delta_X$,
\[ |\psi_t(s,x,\mu)-\psi_t(s',x',\mu)|<\epsilon.\]
\indent\M (F2) Each $\psi_t(s,x,\mu)$ is continuous in $\mu$ at a rate independent of the $(s,x)$ present. That is, for any $\mu\in {\cal P}(S\times X)$ and $\epsilon>0$, there exists $\delta>0$, so that for any $\mu'\in {\cal P}(S\times X)$ satisfying $\rho_{S\times X}(\mu,\mu')<\delta$, as well as any $s\in S$ and $x\in X$,
\[ |\psi_t(s,x,\mu)-\psi_t(s,x,\mu')|<\epsilon.\]
There are a couple of intermediate results, whose proofs are provided in Appendix~\ref{app-a-en2}. Recall that the value functions $v_t$ for an NG are defined around~(\ref{terminal}) and~(\ref{recursive}), while the value functions $v_{nt}$ for finite games are defined around~(\ref{terminal-n}) and~(\ref{recursive-n}). 

\begin{proposition}\label{V-other}
	$v_t(s_t,\sigma_t,x_{[t\bar t]},x_t)$ is continuous in $s_t$ under probabilistically continuous $x_{t'}$'s.
\end{proposition}

\begin{proposition}\label{V-convergence}
	Let $\sigma_t\in {\cal P}(S)$ and $x_{[t\bar t]}\in ({\cal K}(S,G,\gamma,X))^{\bar t-t+1}$ be given. Suppose sequence $s_{t,-1}=(s_{t2},s_{t3},...)$ is sampled from $\sigma_t$, then $v_{nt}(s_{t1},\varepsilon(s^n_{t,-1}),x_{[t\bar t]},x_t)$ will converge to $v_t(s_{t1},\sigma_t,x_{[t\bar t]}$, $x_t)$ in probability at an $s_{t1}$-independent rate, where $s^n_{t,-1}$ stands for the cutoff $(s_{t2},s_{t3},...,s_{tn})$.
	% of the infinite sequence $s_{t,-1}$.
\end{proposition}

Now here comes our main transient result.

\begin{theorem}\label{main}
For state distribution $\sigma_1\in {\cal P}(S)$, suppose $x^*_{[1\bar t]}\equiv(x^*_t)_{t=1,2,...,\bar t}\in ({\cal K}(S,G,\gamma,X))^{\bar t}$ is a probabilistically continuous Markov equilibrium of the nonatomic game $\Gamma(\sigma_1)$. Then, for any $\epsilon>0$ and large enough $n\in \mathbb{N}$, this $x^*_{[1\bar t]}$ is also an $\epsilon$-Markov equilibrium for the game family $(\Gamma_n(\varepsilon(s_1))\mid s_1\in S^n)$ in the sense of $\sigma_{[1\bar t]}\equiv (\sigma_t)_{t=1,..,\bar t}$, where every  $\sigma_t=T_{[1,t-1]}(x^*_{[1,t-1]})\circ\sigma_1$. This means that for any $t=1,...,\bar t$ and $y_t\in {\cal M}(S\times G,X)$,~(\ref{second}) is true:
\[\int_{S^n}v_{nt}\left(s_{t1},\varepsilon(s_{t,-1}),x^*_{[t\bar t]},x^*_t\right)\cdot \sigma^{\;n}_t(ds_t)\geq \int_{S^n}v_{nt}\left(s_{t1},\varepsilon(s_{t,-1}),x^*_{[t\bar t]},y_t\right)\cdot \sigma^{\;n}_t(ds_t)-\epsilon.\]
Furthermore, the same is true in the sense of the stochastic pre-action environment sequence  $\sigma_{n,[1\bar t]}\equiv (\sigma_{nt})_{t=1,...,\bar t}$, where every $\sigma_{nt}$ is a sample over the $\varepsilon(s_t)$'s with $\varepsilon(s_t)=T_{n,[1,t-1]}(x_{[1,t-1]},g_{[1,t-1]},i_{[1,t-1]})\circ \varepsilon(s_1)$, while $(s_1,g_{[1,t-1]},i_{[1,t-1]})$ is distributed according to $(\sigma_1\times \gamma^{t-1}\times \iota^{t-1})^n$. This means that, for any $\epsilon>0$ and large enough $n\in \mathbb{N}$, for any $t=1,...,\bar t$ and $y_t\in {\cal M}(S\times G,X)$,
%\[\int_{S^n}v_{nt}\left(s_{t1},\varepsilon(s_{t,-1}),x^*_{[t\bar t]},x^*_t\right)\cdot \sigma^{\;\;n}_{nt}(ds_t)\geq \int_{S^n}v_{nt}\left(s_{t1},\varepsilon(s_{t,-1}),x^*_{[t\bar t]},y_t\right)\cdot \sigma^{\;\;n}_{nt}(ds_t)-\epsilon.\]
\[\begin{array}{l}
\int_{S^n}\sigma^{\;n}_1(ds_1)\times\int_{G^{n\cdot(t-1)}}\gamma^{n\cdot(t-1)}(dg_{1,t-1]})\times  \int_{I^{n\cdot(t-1)}}\iota^{n\cdot(t-1)}(di_{[1,t-1]})\times v_{nt}\left(s_{t,1},\varepsilon(s_{t,-1}),x^*_{[t\bar t]},x^*_t\right)\\
	\;\;\;\;\;\;\;\;\;\;\;\;\geq \int_{S^n}\sigma^{\;n}_1(ds_1)\times\int_{G^{n\cdot(t-1)}}\gamma^{n\cdot(t-1)}(dg_{1,t-1]})\times \int_{I^{n\cdot(t-1)}}\iota^{n\cdot(t-1)}(di_{[1,t-1]})\times\\
	\;\;\;\;\;\;\;\;\;\;\;\;\;\;\;\;\;\;\;\;\;\;\;\;\;\;\;\;\;\;\;\;\;\;\;\;\times v_{nt}\left(s_{t,1},\varepsilon(s_{t,-1}),x^*_{[t\bar t]},y_t\right)-\epsilon,
\end{array}\]
where both $s_{t,1}$ and $\varepsilon(s_{t,-1})$ come from  $\varepsilon(s_t)$. 
\end{theorem}

%Technical developments of this section can be found in Appendix~\ref{app-a-eq}.
Theorem~\ref{main} says that, when there are enough of them, players in a finite game can agree on an NG equilibrium and expect to lose little on average; also, the distribution based on which ``average'' is taken can be either the NG's state distribution or even an accurate assessment of what players' states would be had they followed the NG equilibrium all along. In the latter option, different players' states can even be correlated. %dependent even though its overall distribution has to remain symmetric. %A prominent feature of an NG equilibrium is its insensitivity to vicissitudes of pre-action environments, which players in a finite game may or may not be observant of.
In the NG limit, the evolution of pre-action environments is deterministic. An equilibrium here, which is necessarily observation-blind to the extent that other players' states and actions do not influence it, serves as a good asymptotic equilibrium for finite games when there are enough players; and, this asymptotic result is independent of the observatory power of players in the finite games. %The root cause of this is, as stated in Theorem~\ref{T-converge}, that the aggregate environment in large games evolves in a nearly deterministic fashion and hence the value of observation has vanished.

\section{A Stationary Situation}\label{stationary}

%\subsection{Model Setup}

Now we study an infinite-horizon model with stationary features. To this end, suppose there is a payoff function $\psi$, so that
\begin{equation}
\psi_t(s,x,\mu)=\alpha^{t-1}\cdot \psi(s,x,\mu),\hspace*{.5in}\forall t=1,2,...,
\end{equation}
where $\alpha\in [0,1)$ is a discount factor.
%Note $\psi$ does not depend on $g$, a point which we shall come back later.
Also, we use $\bar\psi$ for the bound $\bar\psi_1$ that appears in~(\ref{d-payoff}). In addition, suppose there is a state transition function $\theta$, so that
\begin{equation}
\theta_t(s,x,\mu,i)=\theta(s,x,\mu,i),\hspace*{.5in}\forall t=1,2,....
\end{equation}
%Assumptions (S1-g), (S2-g), (F1-g), and (F2-g), are all in place.

For the nonatomic game $\Gamma$ with the above stationary features, we use $x\equiv (x(s,g)\mid s\in S,g\in G)\in {\cal M}(S\times G,X)$ to represent a stationary policy profile. It is a map from the current period's state and pre-action shock to the player's action. Given an $x\in {\cal M}(S\times G,X)$, we denote by $T(x)$ the operator on ${\cal P}(S)$ that converts one state distribution $\sigma$ to its corresponding $T(x)\circ \sigma$ so that following~(\ref{T-def0}), for every $S'\in {\cal B}(S)$, 
\begin{equation}\label{T-def-s}
[T(x)\circ \sigma](S')=\int_S\int_G\int_{I}{\bf 1}\left(\theta(s,x(s,g),M(\sigma,x),i)\in S'\right)\cdot\iota(di)\cdot\gamma(dg)\cdot\sigma(ds).
\end{equation}
An environment $\sigma\in {\cal P}(S)$ is said to be associated with $x$ when
\begin{equation}\label{compatible-s}
\sigma=T(x)\circ \sigma.
\end{equation}
That is, we consider $\sigma\in {\cal P}(S)$ to be associated with $x\in {\cal M}(S\times G,X)$ when the former is invariant under the state transition facilitated by the $T(x)$ operator.

Suppose pre-action environment $\sigma\in {\cal P}(S)$ is associated with policy $x\in {\cal M}(S\times G,X)$. For $t=0,1,...$, we define $v_t(s,\sigma,x,y)$ as the total expected payoff a player can make from period $1$ to $t$, when he starts period 1 with state $s\in S$ and outside environment $\sigma$, while all players keep on using policy $x$ from period $1$ to $t$ with the exception of the current player in the very beginning, who  deviates to $y\in {\cal M}(S\times G,X)$. As a terminal condition, we have
\begin{equation}\label{terminal-s}
v_0(s,\sigma,x,y)=0.
\end{equation}
Due to the stationarity of the setting, we have, for $t=1,2,...$,
\begin{equation}\label{recursive-s}\begin{array}{ll}
v_t(s,\sigma,x,y)=\int_G[\psi(s,y(s,g),M(\sigma,x))\\
\;\;\;\;\;\;\;\;\;\;\;\;\;\;\;\;\;\;\;\;\;\;\;\;+\alpha\cdot\int_I v_{t-1}(\theta(s,y(s,g),M(\sigma,x),i),\sigma,x,x)\cdot \iota(di)]\cdot\gamma(dg).
\end{array}\end{equation}
Using~(\ref{terminal-s}) and~(\ref{recursive-s}), we can inductively show that
\begin{equation}\label{bounded-s}
\mid v_{t+1}(s,\sigma,x,y)-v_t(s,\sigma,x,y)\mid\leq \alpha^t\cdot\bar\psi.
\end{equation}
The sequence $\{v_t(s,\sigma,x,y)\mid t=0,1,...\}$ is thus Cauchy with a limit point $v_\infty(s,\sigma,x,y)$. This $v_\infty(s,\sigma,x,y)$ can be understood as the infinite-horizon total discounted expected payoff a player can obtain by starting with state $s$ and environment $\sigma$, while all players adhere to the action plan $x$ except for the current player in the beginning, who deviates to $y$.

We deem $x^*\in {\cal M}(S\times G,X)$ a Markov equilibrium for the nonatomic game $\Gamma$ when for some $\sigma^*\in {\cal P}(S)$ associated with $x^*$ in the fashion of~(\ref{compatible-s}) and every $y\in {\cal M}(S\times G,X)$,
\begin{equation}\label{first-s}
\int_S v_\infty(s,\sigma^*,x^*,x^*)\cdot \sigma^*(ds)\geq \int_S v_\infty(s,\sigma^*,x^*,y)\cdot \sigma^*(ds).
\end{equation}
Therefore, a policy will be considered an equilibrium when it induces an invariant environment profile under which the policy forms a best response in the long run.

Now we move on to the $n$-player game $\Gamma_n$ with the same stationary features provided by $\psi$, $\theta$, and $\alpha$. Given policy profile $x=(x(s,g)\mid s\in S,g\in G)\in {\cal M}(S\times G,X)$, pre-action shock vector $g=(g_1,...,g_n)\in G^n$, and post-action shock vector $i=(i_1,...,i_n)\in I^n$, we define $T_n(x,g,i)$ as the operator on ${\cal P}_n(S)$ that converts a period's pre-action environment into that of a next period. Following~(\ref{Tn-1-def}), $\varepsilon(s')=T_n(x,g,i)\circ \varepsilon(s)$ is such that
\begin{equation}\label{Tn-1-def-s}
s'_m=\theta\left(s_m,x(s_m,g_m),M_n(s_{-m},g_{-m},x),i_m\right),\hspace*{.5in}\forall m=1,2,...,n.
\end{equation}
Let $v_{nt}(s_1,\varepsilon(s_{-1}),x,y)$ be the total expected payoff player 1 can make from period 1 to $t$, when the player's starting state is $s_1\in S$, other players' initial environments is describable by their aggregate empirical state distribution $\varepsilon(s_{-1})=\varepsilon(s_2,...,s_n)$, and all players adopt the policy $x\in {\cal M}(S\times G,X)$ with the exception that player 1 adopts policy $y\in {\cal M}(S\times G,X)$ in the very beginning. As a terminal condition, we have
\begin{equation}\label{terminal-n-s}
v_{n0}(s_1,\varepsilon(s_{-1}),x,y)=0.
\end{equation}
For $t=1,2,...$, we have that $v_{nt}(s_1,\varepsilon(s_{-1}),x,y)$ equals to
\begin{equation}\label{recursive-n-s}\begin{array}{l}
\int_{G^n}\gamma^n(dg)\times\{\psi\left(s_1,y(s_1,g_1),M_n(s_{-1},g_{-1},x)\right)+\alpha\cdot\int_{I^n}\iota^n(di)\times\\
\;\;\;\;\;\;\;\;\;\;\;\;\times v_{n,t-1}\left(\theta(s_1,y(s_1,g_1),M_n(s_{-1},g_{-1},x),i_1),[T_n(x,g,i)\circ \varepsilon(s)]_{-1},x,x\right)\},
\end{array}\end{equation}
where $[T_n(x,g,i)\circ \varepsilon(s)]_{-1}$ stands for $\varepsilon(s'_{-1})$, such that $\varepsilon(s')=T_n(x,g,i)\circ \varepsilon(s)$. Using~(\ref{terminal-n-s}) and~(\ref{recursive-n-s}), we can inductively show that
\begin{equation}\label{bounded-ss}
\mid v_{n,t+1}(s_1,\varepsilon(s_{-1}),x,y)-v_{nt}(s_1,\varepsilon(s_{-1}),x,y)\mid\leq \alpha^t\cdot\bar\psi.
\end{equation}
Thus, the sequence $\{v_{nt}(s_1,\varepsilon(s_{-1}),x,y)\mid t=0,1,...\}$ is Cauchy with limit  $v_{n\infty}(s_1,\varepsilon(s_{-1}),x,y)$.

We make the following assumptions, which are $t$-independent versions of (S1) to (F2):\\
\indent\M (S1-s) For every $\mu\in {\cal P}(S\times X)$, the function $\theta(\cdot,\cdot,\mu,\cdot)$ is a member of ${\cal K}(S\times X,I,\iota,S)$. \\
\indent\M (S2-s) Not only is it true that $\theta(s,x,\cdot,\cdot)\in {\cal K}({\cal P}(S\times X),I,\iota,S)$ at every $(s,x)\in S\times X$, but the continuity is also achieved at a rate independent of the $(s,x)$ present. \\
\indent\M (F1-s) The function $\psi(s,x,\mu)$ is continuous in $(s,x)$.\\
\indent\M (F2-s) The function $\psi(s,x,\mu)$ is continuous in $\mu$ at an $(s,x)$-independent rate.\\
Here comes our main result for the stationary case.

\begin{theorem}\label{main-s}
	Suppose $x^*\in {\cal K}(S,G,\gamma,X)$ is a probabilistically continuous Markov equilibrium for the nonatomic game $\Gamma$. Let $\sigma^*\in {\cal P}(S)$ be associated with $x^*$ in the fashion of~(\ref{compatible-s}). Then, for any $\epsilon>0$ and large enough $n\in \mathbb{N}$, for any $y\in {\cal M}(S\times G,X)$,
	\[\int_{S^n} v_{n\infty}\left(s_1,\varepsilon(s_{-1}),x^*,x^*\right)\cdot (\sigma^*)^n(ds)\geq \int_{S^n} v_{n\infty}\left(s_1,\varepsilon(s_{-1}),x^*,y\right)\cdot (\sigma^*)^n(ds)-\epsilon.\]
\end{theorem}

Theorem~\ref{main-s} is proved in Appendix~\ref{app-c}. It states that players in a large finite game will not regret much by keeping on adopting a stationary equilibrium for its correspondent nonatomic game. The regret is measured in an average sense, where the underlying invariant state distribution for measuring ``average'' is part of the NG equilibrium. So players can fare well by responding to their individual states in the same fashion indefinitely. %The proof of Theorem~\ref{main-s} leverages on both the existing finite-horizon result Theorem~\ref{main} and the presence of the discount factor $\alpha<1$. The latter ensures that the spotlight need only be trained on payoffs accumulated over the first few periods.

\section{Discussion}\label{discussion}

Using this paper's language and notation, we offer a comparison with the most relevant papers. Within the discrete-time framework while without considering atomic players or players' entries and exits, we have arguably worked with the most general setup. 
%We also touch on the existence of NG equilibria.

%\subsection{Comparison with Most Relevant Literature}

Both Weintraub et al. \cite{WBJvR08} and Weintraub, Benkard, and van Roy \cite{WBvR11} treated competing firms on a common market as players. They allowed for entry and exit of firms, and accounted for the effect of firm density $c$ per unit market size. Roughly speaking, their regular payoff is in the form of $\psi^0(s,c\cdot\mu|_S)-\psi^1(x)$, where $\mu|_S$ stands for the marginal state distribution derivable from the joint state-action distribution $\mu$. Also, firms' state transitions are %isolated as to be
governed by a certain $\theta^0(s,x,i)$ that is independent of the environment $\mu$. %In finite models, $\psi^0$ is replaced by some $\psi^{0\prime}(s,\mu|_S,n,m)$ where $n$ is the number of firms and $m$ the market size. In the limiting regime, it was certainly assumed that
%\begin{equation}
%\lim_{m\rightarrow +\infty}\psi^{0\prime}(s,\mu|_S,n(m),m)=\psi^0(s,c\cdot\mu|_S),
%\end{equation}
%when $\lim_{m\rightarrow +\infty}n(m)/m=c$.

Weintraub et al. \cite{WBJvR08} arrived at something akin to our Theorem~\ref{main}. In the mean time, Weintraub, Benkard, and van Roy \cite{WBvR11} found a stationary policy of the form $x(s)$ to suffice for the NG limit. It was considered oblivious because of firms' abilities to ignore the industry state $c\cdot\mu|_S$. When there are few dominant firms in it, an NG equilibrium was shown to work increasingly well for larger finite models. This is close in spirit to our Theorem~\ref{main-s}. We note that $\theta^0$'s independence of $\mu$ helped greatly with their derivations. While free from the task of dealing with entry, exit, or impacts of market size and number of firms, we have allowed players' state transitions to be profoundly impacted by the environment that their collective states and actions fabricate. Namely, our $\theta_t$ can depend on $\mu$ in virtually arbitrary fashions. %Meanwhile, according to~(\ref{ss-def}) and~(\ref{ss2-def}), such dependence is integral to the pricing game.

Huang, Malhame, and Caines \cite{HMC06} dealt with continuous-time games with the state space $S$ equal to the real line $\Re$. These games' discrete-time counterparts can be obtained by replacing their Brownian motions with symmetric random walks. In particular, we can let the post-action shock space $I$ be $\{-1,+1\}$ and the probability $\iota$ be half on $-1$ and half on $+1$. When thus cast, the earlier work's state transition can be understood as
\begin{equation}
\theta_t(s,x,\mu,i)=\int_{\Re}\theta^0(s,x,s')\cdot \mu|_S(ds')+\bar s^1\cdot i,
\end{equation}
where $\theta^0$ is a function from $\Re\times X\times \Re$ to $\Re$ and $\bar s^1$ is a constant. %; see (1) of \cite{HMC06}.
So there, only the state-distribution portion of the joint state-action distribution $\mu$ of other firms affect the current firm's state transition; its impact is also felt in an average sense; moreover, the effect of the random shock is additive. %This certainly did not cover the case of~(\ref{ss-def}) or~(\ref{ss2-def}).

Their one-period payoff function can be understood as
\begin{equation}
\psi_t(s,x,\mu)=\int_{\Re}\psi^0(s,x,s')\cdot \mu|_S(ds'),
\end{equation}
where $\psi^0$ is a function from $\Re\times X\times \Re$ to $\Re$. %; see (2) of \cite{HMC06}.
%Therefore, payoffs depend on in-action environments also in a special fashion incompatible with~(\ref{ff-def}) and~(\ref{ff2-def}) used in the dynamic pricing game. 
Artificial randomization in decision making turns out to be unnecessary---NG equilibria can be found in the form of $x_t(s)$ rather than the more general $x_t(s,g)$. We, on the other hand, believe that allowing other players' actions to play a role in both state transitions and one-period payoffs can greatly enhance the relevant models' applicabilities. In the competitive pricing situation, for instance, the demand level experienced by a firm is perturbable by prices charged by other firms. It in turn influences not only the firm's present profitability but also its future inventory levels.

%Theorems~\ref{T-converge},~\ref{main} and~\ref{main-s} convey messages similar to, respectively, those carried by Theorems 1, 2, and 4 of Yang \cite{Y15}.
As could be seen from equivalence results such as Aumann \cite{A64b} (Lemma F), using pre-action shocks $g$ and post-action shocks $i$ permit us to effectively deal with both random action plans and random state transitions. These were indeed treated by Yang \cite{Y15} in an alternative transition-probability formulation, with each $\chi_t(s)$ there effectively $x_t(s,\cdot)\circ \gamma^{-1}$ here and each $\tilde g_t(s,x,\mu)$ there effectively $\theta_t(s,x,\mu,\cdot)\circ \iota^{-1}$ here. %It was also shown that pre-action environments of large finite games would converge to those of their NG counterparts, and therefore NG equilibria could be put to good use in large finite games in both transient and stationary situations. 
Due to its need to sample from joint probabilities of the non-product type, however, the earlier work found it necessary to assume discrete state and action spaces. This restriction is removed here through exploitations of the independently generated shocks and tools pertinent to the tightness of probabilities. The latter only requires the current spaces $S$, $G$, and $I$ to be complete. 

We can also apply our results to a dynamic pricing game participated by heterogeneous firms. Since the random demand arrival process is influenced by prices charged by all firms and leftover items are stored for future sales, the finite-player version of this problem is virtually intractable. The usefulness of the transient result Theorem~\ref{main} is thus at full display. To the stationary case also involving production, the stationary result Theorem~\ref{main-s} can further be applied. Moreover, depending on which portion of the outside environment, whether it be merely other firms' prices or both their prices and inventory levels, are observable, there can be different versions of the finite game. The NG approximation renders these differences irrelevant. Details are furnished in Yang \cite{Y17}. 

\section{Concluding Remarks}\label{conclusion}

%Using the weak LLN for empirical distributions, w
We have established links between multi-period Markovian games and their NG counterparts. Our focus is the case where state and action spaces are general metric spaces, and there are independently generated shocks serving as random drivers for decision making and state evolution. In essence, the evolution of player-state distributions in large finite games, though random, resembles in probability the deterministic pathway taken by their NG counterparts. This allows NG equilibria to be well adapted to large finite games.
%In our derivation, transient results played pivotal roles in forming comparable conclusions for stationary systems.
%Our results have been successfully brought to bear on dynamic pricing situations with their own practical significances.

%It is essential in our derivations that both action- and post-action shocks be independent of one another. When there is a global shock commonly experienced by all players, then the dependences of payoffs and transition functions on this shock will translate into inter-dependences among different players. It will not take too much to adapt our approach to this situation. There, the environment pathway of even an NG will not be deterministic. Rather, it will be dependent on the global shocks just experienced.

Still, many dynamic competitive situations not yet covered by existing studies like Huang, Malhame, and Caines \cite{HMC06}  are better described by continuous-time models. These will require vastly different techniques to probe. For one thing, the mathematical induction approach we have taken to deal with multiple periods would not seem to go well with a discrete-time approximation of a continuous-time model. In the latter model, even to identify the environment induced by all players adopting a common policy might involve solving a fixed point problem. Therefore, serious challenges will have to be overcome. 

\newpage

%\noindent{\bf\Large Acknowledgments}

%This research was supported by National Science Foundation Grant CMMI-0854803. The author wishes to thank Professor Thomas G. Kurtz for referring him to Ethier and Kurtz \cite{EK86}, a source instrumental to the proof of Lemma~\ref{joint}.
%\\  \\

\noindent{\bf\Large Appendices}\appendix
\numberwithin{equation}{section}

\section{Technical Developments in Section~\ref{obama}}\label{app-a-en}

Given metric space $A$, the Prohorov metric $\rho_A$ is such that, for any distributions $\pi,\pi'\in {\cal P}(A)$,
\begin{equation}\label{0pm0}
\rho_A(\pi,\pi')=\inf\left(\epsilon>0\mid \pi'((A')^\epsilon)+\epsilon\geq \pi(A'),\;\mbox{ for all }A'\in {\cal B}(A)\right),
\end{equation}
where
\begin{equation}\label{qm0}
(A')^\epsilon=\{a\in A\mid d_A(a,a')<\epsilon\mbox{ for some }a'\in A'\}.
\end{equation}
The metric $\rho_A$ is known to generate the weak topology for ${\cal P}(A)$.\vspace*{.1in}

According to Parthasarathy \cite{P05} (Theorem II.7.1), the strong LLN applies to the empirical distribution under the weak topology, and hence under the Prohorov metric. In the following, we state its weak version.

\begin{lemma}\label{p-Prohorov}
	Given separable metric spaces $A$ and $B$, suppose distribution $\pi_A\in {\cal P}(A)$ and measurable mapping $y\in {\cal M}(A,B)$. Then, for any $\epsilon>0$, as long as $n$ is large enough,
	\[ (\pi_A)^n\left(\left\{a=(a_1,...,a_n)\in A^n\mid \rho_B(\varepsilon(a) \cdot y^{-1},\pi \cdot y^{-1})<\epsilon\right\}\right)>1-\epsilon. \]
\end{lemma}

For separable metric space $A$, point $a\in A$, and the $(n-1)$-point empirical distribution space $\pi\in {\cal P}_{n-1}(A)$, we use $(a,\pi)_n$ to represent the member of ${\cal P}_n(A)$ that has an additional $1/n$ weight on the point $a$, but with probability masses in $\pi$ being reduced to $(n-1)/n$ times of their original values. For $a\in A^n$ and $m=1,...,n$, we have $(a_m,\varepsilon(a_{-m}))_n=\varepsilon(a)$. Concerning the Prohorov metric, we have also a simple but useful observation.

\begin{lemma}\label{p-mc8}
	Let $A$ be a separable metric space. Then, for any $n=2,3,...$, $a\in A$, and $\pi\in {\cal P}_{n-1}(A)$,
	\[\rho_A((a,\pi)_n,\pi)\leq \frac{1}{n}. \]
\end{lemma}
\noindent{\bf Proof: }% of Lemma~\ref{p-mc8}: }
Let $A'\in {\cal B}(A)$ be chosen. If $a\notin A'$, then
\begin{equation}
(a,\pi)_n(A')\leq \pi(A')\leq (a,\pi)_n(A')+\frac{1}{n};
\end{equation}
if $a\in A'$, then
\begin{equation}
(a,\pi)_n(A')-\frac{1}{n}\leq \pi(A')\leq (a,\pi)_n(A').
\end{equation}
Hence, it is always true that
\begin{equation}
\mid (a,\pi)_n(A')-\pi(A')\mid\leq \frac{1}{n}.
\end{equation}
In view of~(\ref{0pm0}) and~(\ref{qm0}), we have
\begin{equation}
\rho_A\left((a,\pi)_n,\pi\right)\leq \frac{1}{n}.
\end{equation}
We have thus completed the proof.\qed

The following result is important for showing the near-trajectory evolution of aggregate environments in large multi-period games.

\begin{lemma}\label{joint}
Given separable metric space $A$ and complete separable metric spaces $B$ and $C$, suppose $y_n\in {\cal M}(A^n,B^n)$ for every $n\in \mathbb{N}$, $\pi_A\in {\cal P}(A)$, $\pi_B\in {\cal P}(B)$, and $\pi_C\in {\cal P}(C)$. If
\[(\pi_A)^n\left(\{a\in A^n\mid \rho_B(\varepsilon(y_n(a)),\pi_B)<\epsilon\}\right)>1-\epsilon, \]
for any $\epsilon>0$ and any $n$ large enough, then
\[ (\pi_A\times\pi_C)^n\left(\{(a,c)\in (A\times C)^n\mid\rho_{B\times C}(\varepsilon(y_n(a),c),\pi_B\times\pi_C)<\epsilon\}\right)>1-\epsilon, \]
for any $\epsilon>0$ and any $n$ large enough.
\end{lemma}
\noindent{\bf Proof: }% of Lemma~\ref{joint}: }
Suppose sequence $\{\pi'_{B1},\pi'_{B2},...\}$ weakly converges to the given probability measure $\pi_B$, and sequence $\{\pi'_{C1},\pi'_{C2},...\}$ weakly converges to the given probability measure $\pi_C$. We are to show that the sequence $\{\pi'_{B1}\times \pi'_{C1},\pi'_{B2}\times\pi'_{C2},...\}$ weakly converges to $\pi_B\times \pi_C$.

Let $F(B)$ denote the family of uniformly continuous real-valued functions on $B$ with bounded support. Let $F(C)$ be similarly defined for $C$. We certainly have
\begin{equation}\label{abo}\left\{\begin{array}{l}
\lim_{k\rightarrow +\infty}\int_B f(b)\cdot\pi'_{Bk}(db)=\int_B f(b)\cdot\pi_B(db),\hspace*{.5in}\forall f\in F(B),\\
\lim_{k\rightarrow +\infty}\int_C f(c)\cdot\pi'_{Ck}(dc)=\int_C f(c)\cdot\pi_C(dc),\hspace*{.5in}\forall f\in F(C).
\end{array}\right.\end{equation}
Define $F$ so that
\begin{equation}\label{bel}\begin{array}{l}
F=\{f\mid f(b,c)=f_B(b)\cdot f_C(c)\;\mbox{ for any }(b,c)\in B\times C,\\
\;\;\;\;\;\;\;\;\;\;\;\;\mbox{ where }f_B\in F(B)\cup\{{\bf 1}\}\mbox{ and }f_C\in F(C)\cup\{{\bf 1}\}\},
\end{array}\end{equation}
where ${\bf 1}$ stands for the function whose value is 1 everywhere. By~(\ref{abo}) and~(\ref{bel}),
\begin{equation}\label{haha}
\lim_{k\rightarrow +\infty}\int_{B\times C}f(b,c)\cdot (\pi'_{Bk}\times \pi'_{Ck})(d(b,c))=\int_{B\times C}f(b,c)\cdot (\pi_B\times \pi_C)(d(b,c)).
\end{equation}

According to Ethier and Kurtz \cite{EK86} (Proposition III.4.4), $F(B)$ and $F(C)$ happen to be ${\cal P}(B)$ and ${\cal P}(C)$'s convergence determining families, respectively. As $B$ and $C$ are complete, Ethier and Kurtz (\cite{EK86}, Proposition III.4.6, whose proof involves Prohorov's Theorem, i.e., the equivalence between tightness and relative compactness of a collection of probability measures defined for complete separable metric spaces) further states that $F$ as defined through~(\ref{bel}) is convergence determining for ${\cal P}(B\times C)$. Therefore, we have the desired weak convergence by~(\ref{haha}).

Let $\epsilon>0$ be given. In view of the above product-measure convergence and the equivalence between the weak topology and that induced by the Prohorov metric, there must be $\delta_B>0$ and $\delta_C>0$, such that $\rho_B(\pi'_B,\pi_B)<\delta_B$ and $\rho_C(\pi'_C,\pi_C)<\delta_C$ will imply
\begin{equation}\label{above}
(\rho_B\times\rho_C)(\pi'_B\times\pi'_C,\pi_B\times\pi_C)<\epsilon.
\end{equation}

By~(\ref{0pm0}) and the given hypothesis, there is $\bar n^1\in \mathbb{N}$, so that for $n=\bar n^1,\bar n^1+1,...$,
\begin{equation}\label{ab-f}
(\pi_A)^n(\tilde A_n)>1-\frac{\epsilon}{2},
\end{equation}
where $\tilde A_n$ contains all $a\in A^n$ such that
\begin{equation}\label{deltab}
\rho_B(\varepsilon(y_n(a)),\pi_B)<\delta_B.
\end{equation}
By~(\ref{0pm0}) and Lemma~\ref{p-Prohorov}, on the other hand, there is $\bar n^2\in \mathbb{N}$, so that for $n=\bar n^2,\bar n^2+1,...$,
\begin{equation}\label{c-f}
(\pi_C)^n(\tilde C_n)>1-\frac{\epsilon}{2},
\end{equation}
where $\tilde C_n$ contains all $c\in C^n$ such that
\begin{equation}\label{deltac}
\rho_C(\varepsilon(c),\pi_C)<\delta_C.
\end{equation}
For any $n=\bar n^1\vee\bar n^2,\bar n^1\vee\bar n^2+1,...$, let $(a,c)$ be an arbitrary member of $\tilde A_n\times\tilde C_n$. We have from~(\ref{above}),~(\ref{deltab}), and~(\ref{deltac}) that,
\begin{equation}
(\rho_B\times\rho_C)(\varepsilon(y_n(a),c),\pi_B\times \pi_C)<\epsilon.
\end{equation}
Noting the facilitating $(a,c)$ is but an arbitrary member of $\tilde A_n\times\tilde C_n$, we see that
\begin{equation}\begin{array}{l}
(\pi_A\times \pi_C)^n\left(\{(a,c)\in (A\times C)^n\mid\rho_{B\times C}(\varepsilon(y_n(a),c),\pi_B\times\pi_C)<\epsilon\}\right)\\
\;\;\;\;\;\;\;\;\;\;\;\;\;\;\;\;\;\;\;\;\;\;\;\;\geq (\pi_A)^n(\tilde A_n)\times (\pi_C)^n(\tilde C_n),
\end{array}\end{equation}
which by~(\ref{ab-f}) and~(\ref{c-f}), is greater than $1-\epsilon$. \qed

Because the equivalence between tightness and relative compactness of a collection of probability measures is indirectly related to the proof of Lemma~\ref{joint}, we require $B$ and $C$ to be complete separable metric spaces. %This is not stringent, as closed real intervals endowed with the Euclidean metric, as well as finite and countable sets endowed with the discrete metric, where every two different points has a distance 1, are all complete separable metric spaces.

%For any separable metric spaces $A$, $B$, and $C$, and distribution $\pi_A\in {\cal P}(A)$, we define $Q(A,B,C,\pi_A)\subseteq {\cal M}(A\times B,C)$, to the effect that any $y\in Q(A,B,C,\pi_A)$ is uniformly continuous, as a $B$-to-random-$C$ function, in the following $\pi_A$-defined probabilistic sense: for any $\epsilon>0$, there exists $\delta>0$, so that for any $b,b'\in B$ satisfying $d_B(b,b')<\delta$,
%\[ \pi_A(\{a\in A\mid d_C(y(a,b),y(a,b'))<\epsilon\})>1-\epsilon. \]
%When $y\in {\cal M}(A\times B,C)$ is uniformly continuous as a function of $b\in B$ for $\pi_A$-almost every $a\in A$, it will certainly belong to $Q(A,B,C,\pi_A)$. The following result uses this newly defined concept.
%Suppose $y\in {\cal M}(A\times B,C)$ is given. For each $b\in B$, we may define $\pi_A\cdot y^{-1}(b)\in {\cal P}(C)$, so that for any $C'\in {\cal B}(C)$,
%\[ [\pi_A\cdot y^{-1}(b)](C')=\pi_A(y^{-1}(b)(C'))=\pi_A(\{a\in A\mid y(a,b)\in C'\}).\]
%It may be seen that $y\in Q(A,B,C,\pi_A)$ will lead to the continuity of $\pi_A\cdot y^{-1}(\cdot)$ as a map from $B$ to ${\cal P}(C)$. As it is not directly used, we omit the proof of this result here.

\begin{lemma}\label{cojoint}
	Given separable metric spaces $A$, $B$, $C$, and $D$, as well as distributions $\pi_A\in {\cal P}(A)$, $\pi_B\in {\cal P}(B)$, and $\pi_C\in {\cal P}(C)$, suppose $y_n\in {\cal M}(A^n,B^n)$ for every $n\in \mathbb{N}$ and $z\in {\cal K}(B,C,\pi_C,D)$. If
	\[ (\pi_A\times\pi_C)^n\left(\{a\in A^n,c\in C^n\mid \rho_{B\times C}(\varepsilon(y_n(a),c),\pi_B\times \pi_C)<\epsilon\}\right)>1-\epsilon, \]
	for any $\epsilon>0$ and any $n$ large enough, then
	\[ (\pi_A\times\pi_C)^n\left(\left\{a\in A^n,c\in C^n\mid \rho_D(\varepsilon(y_n(a),c)\cdot z^{-1},(\pi_B\times \pi_C)\cdot z^{-1})<\epsilon\right\}\right)>1-\epsilon, \]
	for any $\epsilon>0$ and any $n$ large enough.
\end{lemma}
\noindent{\bf Proof: }% of Lemma~\ref{cojoint}: }
Let $\epsilon>0$ be given. Since $z\in {\cal K}(B,C,\pi_C,D)$, there exist $C'\in {\cal B}(C)$ satisfying
\begin{equation}\label{uuu}
\pi_C(C')>1-\frac{\epsilon}{2},
\end{equation}
as well as
\begin{equation}\label{sd}
\delta\in (0,\epsilon/2],
\end{equation}
such that for any $b,b'\in B$ satisfying $d_B(b,b')<\delta$ and any $c\in C'$,
\begin{equation}
d_D(z(b,c),z(b',c))<\epsilon.
\end{equation}
For any subset $D'$ in ${\cal B}(D)$, we therefore have
\begin{equation}
(z^{-1}(D'))^\delta\cap (B\times C')\subseteq z^{-1}((D')^\epsilon).
\end{equation}
This leads to $(z^{-1}(D'))^\delta\setminus (B\times (C\setminus C'))\subseteq z^{-1}((D')^\epsilon)$, and hence due to~(\ref{uuu}),
\begin{equation}\label{fact10}
(\pi_B\times \pi_C)\left(z^{-1}((D')^\epsilon)\right)\geq (\pi_B\times \pi_C)\left((z^{-1}(D'))^\delta\right)-\frac{\epsilon}{2}.
\end{equation}

On the other hand, by the hypothesis, we know for $n$ large enough,
\begin{equation}\label{abo2}
(\pi_A\times\pi_C)^n(E'_n)>1-\delta,
\end{equation}
where
\begin{equation}\label{abo3}
E'_n=\{a\in A^n,c\in C^n\mid \rho_{B\times C}(\varepsilon(y_n(a),c),\pi_B\times \pi_C)<\delta\}\in {\cal B}^n(A\times C).
\end{equation}
By~(\ref{abo3}), for any $(a,b)\in E'_n$ and $F'\in {\cal B}(B\times C)$,
\begin{equation}\label{fact20}
(\pi_B\times \pi_C)((F')^\delta)\geq [\varepsilon(y_n(a),c)](F')-\delta.
\end{equation}
%Though closed subsets are used in the definition~(\ref{0pm0}), here for $\rho_{B\times C}$, we have opted for the version without the closedness requirement.

Combining the above, we have, for any $(a,c)\in E'_n$ and $D'\in {\cal B}(D)$,
\begin{equation}\begin{array}{l}
[(\pi_B\times \pi_C)\cdot z^{-1}]((D')^\epsilon)=(\pi_B\times \pi_C)(z^{-1}((D')^\epsilon))\\
\;\;\;\;\;\;\geq (\pi_B\times\pi_C)((z^{-1}(D'))^\delta)-\epsilon/2\geq [\varepsilon(y_n(a),c)](z^{-1}(D'))-\delta-\epsilon/2\\
\;\;\;\;\;\;\geq [\varepsilon(y_n(a),c)](z^{-1}(D'))-\epsilon=([\varepsilon(y_n(a),c)]\cdot z^{-1})(D')-\epsilon.
\end{array}\end{equation}
where the first inequality is due to~(\ref{fact10}), the second inequality is due to~(\ref{fact20}), and the third inequality is due to~(\ref{sd}). That is, we have
\begin{equation}
\rho_D\left(\varepsilon(y_n(a),c)\cdot z^{-1},(\pi_B\times\pi_C)\cdot z^{-1}\right)\leq\epsilon,\hspace*{.5in}\forall (a,c)\in E'_n.
\end{equation}
In view of~(\ref{sd}) and~(\ref{abo2}), we have the desired result. \qed

We can now prove Proposition~\ref{T-onestep} and then Theorem~\ref{T-converge}.

\noindent{\bf Proof of Proposition~\ref{T-onestep}: }Let $t=1,...,\bar t-1$ and $x\in {\cal K}(S,G,\gamma,X)$ be given. Define map $z\in {\cal M}(S\times G\times I,S)$, so that
\begin{equation}\label{z-def}
z(s,g,i)=\theta_t\left(s,x(s,g),M(\sigma,x),i\right),\hspace*{.5in}\forall s\in S,g\in G,i\in I.
\end{equation}
In view of~(\ref{T-def}) and~(\ref{z-def}), we have, for any $S'\in{\cal B}(S)$,
\begin{equation}\label{message1}\begin{array}{l}
[T_t(x)\circ \sigma](S')=\int_S\int_G\int_I{\bf 1}(z(s,g,i)\in S')\cdot\iota(di)\cdot\gamma(dg)\cdot\sigma(ds)\\
\;\;\;\;\;\;=(\sigma\times\gamma\times\iota)(\{(s,g,i)\in S\times G\times I\mid z(s,g,i)\in S'\})=(\sigma\times\gamma\times\iota)(z^{-1}(S')).
\end{array}\end{equation}
For $n\in \mathbb{N}$, $g=(g_1,...,g_n)\in G^n$, and $i=(i_1,...,i_n)\in I^n$, also define operator $T'_n(g,i)$ on ${\cal P}_n(S)$ so that $T'_n(g,i)\circ \varepsilon(s)=\varepsilon(s')$, where for $m=1,2,...,n$,
\begin{equation}\label{T'-def}
s'_m=z(s_m,g_m,i_m)=\theta_t\left(s_m,x(s_m,g_m),M(\sigma,x),i_m\right).
\end{equation}
It is worth noting that~(\ref{T'-def}) is different from the earlier~(\ref{Tn-1-def}). In view of~(\ref{z-def}) and~(\ref{T'-def}), we have, for $S'\in {\cal B}(S)$, that $[T'_n(g,i)\circ \varepsilon(s)](S')$ equals
\begin{equation}\label{message2}
\frac{1}{n}\cdot\sum_{m=1}^n{\bf 1}\left(z(s_m,g_m,i_m)\in S'\right)=\varepsilon((s_1,g_1,i_1),...,(s_n,g_n,i_n))\left(z^{-1}(S')\right).
\end{equation}
Combining~(\ref{message1}) and~(\ref{message2}), we arrive to a key observation that
\begin{equation}\label{translation}
T_t(x)\circ\sigma=(\sigma\times\gamma\times\iota)\cdot z^{-1},\hspace*{.2in}\mbox{ while }\hspace*{.1in}T'_n(g,i)\circ\varepsilon(s)=\varepsilon(s,g,i)\cdot z^{-1}.
\end{equation}
In the rest of the proof, we first show the asymptotic closeness between $T_t(x)\circ\sigma$ and $T'_n(g,i)\circ \varepsilon(s_n(a))$, and then that between the latter and $T_{nt}(x,g,i)\circ \varepsilon(s_n(a))$.

First, due to the hypothesis on the convergence of $\varepsilon(s_n(a))$ to $\sigma$, the completeness of the spaces $S$, $G$, and $I$ and hence also the completeness of $G\times I$, as well as Lemma~\ref{joint},
\begin{equation}\label{okk}
(\pi\times\gamma\times\iota)^n(\{(a,g,i)\in (A\times G\times I)^n\mid \rho_{S\times G\times I}(\varepsilon(s_n(a),g,i),\sigma\times\gamma\times\iota)<\epsilon'\})>1-\epsilon',
\end{equation}
for any $\epsilon'>0$ and any $n$ large enough. By (S1) and the fact that $x\in {\cal K}(S,G,\gamma,X)$, we may see that $z$ as defined through~(\ref{z-def}) is a member of ${\cal K}(S,G\times I,\gamma\times\iota,S)$. By Lemma~\ref{cojoint}, this fact along with~(\ref{okk}) will lead to the strict dominance of $1-\epsilon'$ by
\begin{equation}
(\pi\times\gamma\times\iota)^n(\{(a,g,i)\in (A\times G\times I)^n\mid \rho_S(\varepsilon(s_n(a),g,i)\cdot z^{-1},(\sigma\times\gamma\times\iota)\cdot z^{-1})<\epsilon'\}),
\end{equation}
for any $\epsilon'>0$ and any $n$ large enough. By~(\ref{translation}), this is equivalent to that, given $\epsilon>0$, there exists $\bar n^1\in \mathbb{N}$ so that for any $n=\bar n^1,\bar n^1+1,...$,
\begin{equation}\label{a-ineq}
(\pi\times\gamma\times\iota)^n\left(\tilde A_n(\epsilon)\right)>1-\frac{\epsilon}{2},
\end{equation}
where $\tilde A_n(\epsilon)\in {\cal B}^n(A\times G\times I)$ is equal to
\begin{equation}\label{aa-def}
\left\{(a,g,i)\in (A\times G\times I)^n\mid \rho_S\left(T_t(x)\circ\sigma,T'_n(g,i)\circ \varepsilon(s_n(a))\right)<\frac{\epsilon}{2}\right\}.
\end{equation}

Next, note that the only difference between $T_{nt}(x,g,i)\circ \varepsilon(s_n(a))$ and $T'_n(g,i)\circ \varepsilon(s_n(a))$ lies in that $\varepsilon(s_{n,-m}(a),g_{-m})$ is used in the former as in~(\ref{Tn-1-def}) whereas $\sigma\times\gamma$ is used in the latter as in~(\ref{T'-def}). Here, $s_{n,-m}(a)$ refers to the vector $(s_{n1}(a),...,s_{n,m-1}(a),s_{n,m+1}(a),...,s_{nn}(a))$. By (S2), there is $\delta\in (0,\epsilon/4]$ and $I'\in {\cal B}(I)$ with
\begin{equation}\label{comeon}
\iota(I')>1-\frac{\epsilon}{4},
\end{equation}
so that for any $(s,g,i)\in S\times G\times I'$ and any $\mu'\in {\cal P}(S\times X)$ satisfying $\rho_{S\times X}(M(\sigma,x),\mu')<\delta$,
\begin{equation}\label{great}
d_S\left(\theta_t(s,x(s,g),M(\sigma,x),i),\theta_t(s,x(s,g),\mu',i)\right)<\frac{\epsilon}{2}.
\end{equation}
For each $n\in \mathbb{N}$, define $I'_n$ so that
\begin{equation}\label{def-iprime}
I'_n=\left\{i=(i_1,...,i_n)\in I^n\mid \mbox{more than }\left(1-\frac{\epsilon}{2}\right)\cdot n\mbox{ components come from }I'\right\}.
\end{equation}
Also important is that by~(\ref{great}) and~(\ref{def-iprime}), for any $S'\in {\cal B}(S)$ and $i=(i_1,...,i_n)\in I'_n$,
\begin{equation}\label{key1}
\left[T_{nt}(x,g,i)\circ \varepsilon(s_n(a))\right]\left((S')^{\epsilon/2}\right)+\frac{\epsilon}{2}\geq \left[T'_n(g,i)\circ \varepsilon(s_n(a))\right](S'),
\end{equation}
whenever
\begin{equation}\label{likely}
\rho_{S\times X}\left(M(\sigma,x),M_n(s_{n,-m}(a),g_{-m},x)\right)<\delta.
\end{equation}

It can be shown that $I'_n$ will occupy a big chunk of $I^n$ as measured by $\iota^n$ when $n$ is large. Define map $q$ from $I$ to $\{0,1\}$ so that $q(i)=1$ or 0 depending on whether or not $i\in I'$. By~(\ref{comeon}), $\iota\cdot q^{-1}$ is a Bernoulli distribution with $(\iota\cdot q^{-1})(\{1\})>1-\epsilon/4$. So by~(\ref{def-iprime}), $I'_n$ contains all $i=(i_1,...,i_n)\in I^n$ that satisfy
\begin{equation}
\rho_{\{0,1\}}(\varepsilon(i)\cdot q^{-1},\iota\cdot q^{-1})<\frac{\epsilon}{4}.
\end{equation}
Therefore, by Lemma~\ref{p-Prohorov}, there exits $\bar n^2\in \mathbb{N}$, so that for $n=\bar n^2,\bar n^2+1,...$,
\begin{equation}\label{iota-i}
\iota^n(I'_n)>1-\frac{\epsilon}{4}.
\end{equation}
We can also demonstrate that~(\ref{likely}) will be highly likely when $n$ is large. By Lemma~\ref{joint} and the hypothesis on the convergence of $\varepsilon(s_n(a))$ to $\sigma$, we know $\varepsilon(s_n(a),g)$ will converge to $\sigma\times \gamma$ in probability. Due to Lemma~\ref{p-mc8}, this conclusion applies to the sequence $\varepsilon(s_{n,-m}(a),g_{-m})$ as well. The fact that $x\in {\cal K}(S,G,\gamma,X)$ certainly leads to $(\mbox{prj}_S,x)\in {\cal K}(S,G,\gamma,S\times X)$. So by Lemma~\ref{cojoint}, there is $\bar n^3\in \mathbb{N}$, so that for $n=\bar n^3,\bar n^3+1,...$,
\begin{equation}\label{b-ineq}
(\pi^n\times \gamma^n)\left(\tilde B_n(\delta)\right)>1-\frac{\epsilon}{4},
\end{equation}
where
\begin{equation}\label{bb-def}
\tilde B_n(\delta)=\{(a,g)\in A^n\times G^n\mid\mbox{(\ref{likely}) is true}\}\in {\cal B}^n(A\times G).
\end{equation}
Consider arbitrary $n=\bar n^1\vee\bar n^2\vee\bar n^3,\bar n^1\vee\bar n^2\vee\bar n^3+1,...$, $(a,g,i)\in \tilde A_n(\epsilon)\cap(\tilde B_n(\delta)\times I'_n)$, and $S'\in {\cal B}(S)$. By~(\ref{0pm0}) and~(\ref{aa-def}), we see that
\begin{equation}
[T'_n(g,i)\circ \varepsilon(s_n(a))]\left((S')^{\epsilon/2}\right)+\frac{\epsilon}{2}\geq [T_t(x)\circ\sigma](S').
\end{equation}
Combining this with~(\ref{key1}),~(\ref{likely}), and~(\ref{bb-def}), we obtain
\begin{equation}
[T_{nt}(x,g,i)\circ \varepsilon(s_n(a))]\left((S')^\epsilon\right)+\epsilon\geq [T'_n(g,i)\circ \varepsilon(s_n(a))]\left((S')^{\epsilon/2}\right)+\frac{\epsilon}{2}\geq [T_t(x)\circ\sigma](S').
\end{equation}
According to~(\ref{0pm0}), this means
\begin{equation}
\rho_S\left(T_{nt}(x,g,i)\circ \varepsilon(s_n(a)),T_t(x)\circ\sigma\right)\leq\epsilon.
\end{equation}
Therefore, for $n\geq \bar n^1\vee\bar n^2\vee\bar n^3$,
\begin{equation}\begin{array}{l}
(\pi\times\gamma\times\iota)^n\left(\{(a,g,i)\in (A\times G\times I)^n\mid  \rho_S(T_{nt}(x,g,i)\circ \varepsilon(s_n(a)),T_t(x)\circ\sigma)\leq\epsilon\}\right)\\
\;\;\;\;\;\;\;\;\;\;\;\;\;\;\;\;\;\;\;\;\;\;\;\;\geq (\pi\times\gamma\times\iota)^n\left(\tilde A_n(\epsilon)\cap(\tilde B_n(\delta)\times I'_n)\right),
\end{array}\end{equation}
whereas the latter is, in view of~(\ref{a-ineq}),~(\ref{iota-i}), and~(\ref{b-ineq}), greater than $1-\epsilon$. \qed

\noindent{\bf Proof of Theorem~\ref{T-converge}: }We use induction to show that, for each $\tau=0,1,...,\bar t-t+1$,
\begin{equation}\label{att}
\left(\sigma_t\times\gamma^\tau\times\iota^\tau\right)^n\left(\tilde A_{n\tau}(\epsilon)\right)>1-\frac{\epsilon}{\bar t-t+2},
\end{equation}
for any $\epsilon>0$ and $n$ large enough, where $\tilde A_{n\tau}(\epsilon)\in {\cal B}^n(S\times G^\tau\times I^\tau)$ is such that, for any $(s_t,g_{[t,t+\tau-1]},i_{[t,t+\tau-1]})\in \tilde A_{n\tau}(\epsilon)$,
\begin{equation}
\rho_S\left(T_{n,[t,t+\tau-1]}(x_{[t,t+\tau-1]},g_{[t,t+\tau-1]},i_{[t,t+\tau-1]})\circ \varepsilon(s_t),T_{[t,t+\tau-1]}(x_{[t,t+\tau-1]})\circ\sigma_t\right)<\epsilon.
\end{equation}
Once the above is achieved, we can then define $\tilde A_n(\epsilon)$ required in the theorem by
\begin{equation}
\tilde A_n(\epsilon)=\bigcap_{\tau=0}^{\bar t-t+1}\left[\tilde A_{n\tau}(\epsilon)\times G^{n\cdot(\bar t-t+1-\tau)}\times I^{n\cdot (\bar t-t+1-\tau)}\right].
\end{equation}
This and~(\ref{att}) will lead to
\begin{equation}
\left(\sigma_t\times\gamma^{\bar t-t+1}\times\iota^{\bar t-t+1}\right)^n\left(\tilde A_n(\epsilon)\right)> \left(1-\frac{\epsilon}{\bar t-t+2}\right)^{\bar t-t+2}>1-\epsilon,
\end{equation}
for any $\epsilon>0$ and $n$ large enough.

Now we proceed with the induction process. First, note that $T_{n,[t,t-1]}\circ \varepsilon(s_t)$ is merely $\varepsilon(s_t)$ itself and $T_{[t,t-1]}\circ\sigma_t$ is merely $\sigma_t$ itself. Hence, we will have~(\ref{att}) for $\tau=0$ for any $\epsilon>0$ and $n$ large enough just by Lemma~\ref{p-Prohorov}. Then, for some $\tau=1,2,...,\bar t-t+1$, suppose
\begin{equation}
\left(\sigma_t\times\gamma^{\tau-1}\times\iota^{\tau-1}\right)^n\left(\tilde A_{n,\tau-1}(\epsilon)\right)>1-\frac{\epsilon}{\bar t-t+2},
\end{equation}
for any $\epsilon>0$ and $n$ large enough. We may apply Proposition~\ref{T-onestep} to the above, while at the same time identifying $S\times G^{\tau-1}\times I^{\tau-1}$ with $A$, $\sigma_t\times\gamma^{\tau-1}\times\iota^{\tau-1}$ with $\pi$, $x_{t+\tau-1}$ with $x$, $T_{n,[t,t+\tau-2]}(x_{[t,t+\tau-2]},g_{[t,t+\tau-2]},i_{[t,t+\tau-2]})\circ \varepsilon(s_t)$ with $\varepsilon(s_n(a))$, and $T_{[t,t+\tau-2]}(x_{[t,t+\tau-2]})\circ\sigma_t$ with $\sigma$. This way, we will verify~(\ref{att}) for any $\epsilon>0$ and $n$ large enough. Therefore, the induction process can be completed. \qed

\section{Technical Developments in Section~\ref{eq-converge}}\label{app-a-en2}

\noindent{\bf Proof of Proposition~\ref{V-other}: }Because payoff functions are bounded, the value functions are bounded too. We then prove by induction on $t$. By~(\ref{terminal}), we know the result is true for $t=\bar t+1$. Suppose for some $t=\bar t,\bar t-1,...,2$, we have the continuity of $v_{t+1}(s_{t+1},\sigma_{t+1},x_{[t+1,\bar t]},x_{t+1})$ in $s_{t+1}$. By this induction hypothesis, the probabilistic continuity of $x_t$, (S1), (F1), and the boundedness of the value functions, we see the continuity of the right-hand side of~(\ref{recursive}) in $s_t$. So, $v_t(s_t,\sigma_t,x_{[t\bar t]},x_t)$ is continuous in $s_t$, and we have completed our induction process. \qed

\noindent{\bf Proof of Proposition~\ref{V-convergence}: }We prove by induction on $t$. By~(\ref{terminal}) and~(\ref{terminal-n}), we know the result is true for $t=\bar t+1$. Suppose for some $t=\bar t,\bar t-1,...,2$, we have the convergence of $v_{n,t+1}(s_{t+1,1},\varepsilon(s^n_{t+1,-1}),x_{[t+1,\bar t]},x_{t+1})$ to $v_{t+1}(s_{t+1,1},\sigma_{t+1},x_{[t+1,\bar t]},x_{t+1})$ at an $s_{t+1,1}$-independent rate when $s_{t+1,-1}=(s_{t+1,2},s_{t+1,3},...)$ is sampled from $\sigma_{t+1}$. Now, suppose $s_{t,-1}=(s_{t2},s_{t3},...)$ is sampled from $\sigma_t$. Let also $g=(g_1,g_2,...)$ be generated through sampling on $(G,{\cal B}(G),\gamma)$ and $i=(i_1,i_2,...)$ be generated through sampling on $(I,{\cal B}(I),\iota)$. In the remainder of the proof, we let $s^n_t=(s_{t1},s_{t2},...,s_{tn})$ for any arbitrary $s_{t1}\in S$, $g^n=(g_1,...,g_n)$ and $i^n=(i_1,...,i_n)$.

Due to Lemma~\ref{p-Prohorov}, $\varepsilon(s^n_{t,-1})$ will converge to $\sigma_t$. By Lemma~\ref{p-mc8}, $\varepsilon(s^n_t)$ will converge to $\sigma_t$ at an $s_{t1}$-independent rate. By Proposition~\ref{T-onestep}, we know that $T_{nt}(x_t,g^n,i^n)\circ \varepsilon(s^n_t)$ will converge to $T_t(x_t)\circ\sigma_t$ in probability at an $s_{t1}$-independent rate, and by Lemma~\ref{p-mc8} again, so will $[T_{nt}(x_t,g^n,i^n)\circ \varepsilon(s^n_t)]_{-1}$ to $T_t(x_t)\circ\sigma_t$. Now Lemma~\ref{joint} will lead to the convergence in probability of $\varepsilon(s^n_{t,-1},g^n_{-1})$ to $\sigma_t\times\gamma$. Due to $x_t$'s probabilistic continuity, Lemma~\ref{cojoint} will lead to the convergence in probability of $M_n(s^n_{t,-1},g^n_{-1},x_t)$ to $M(\sigma_t,x_t)$. Thus,\\
\indent\M 1. $\psi_t(s_{t1},x_t(s_{t1},g_1),M_n(s^n_{t,-1},g^n_{-1},x_t))$ will converge to $\psi_t(s_{t1},x_t(s_{t1},g_1),M(\sigma_t,x_t))$ in probability at an $s_{t1}$-independent rate due to (F2); \\
\indent\M 2. $v_{n,t+1}(\theta_t(s_{t1},x_t(s_{t1},g_1),M_n(s^n_{t,-1},g^n_{-1},x_t),i_1),[T_{nt}(x_t,g^n,i^n)\circ \varepsilon(s^n_t)]_{-1},x_{[t+1,\bar t]},x_{t+1})$ will converge to $v_{t+1}(\theta_t(s_{t1},x_t(s_{t1},g_1),M_n(s^n_{t,-1},g^n_{-1}),x_t),i_1),T_t(x_t)\circ \sigma_t,x_{[t+1,\bar t]},x_{t+1})$ in probability at an $s_{t1}$-independent rate due to the induction hypothesis; the latter will in turn converge to $v_{t+1}(\theta_t(s_{t1},x_t(s_{t1},g_1),M(\sigma_t,x_t),i_1),T(x_t)\circ\sigma_t,x_{[t+1,\bar t]},x_{t+1})$ in probability at an $s_{t1}$-independent rate due to (S2) and Proposition~\ref{V-other}.

As per-period payoffs are bounded, all value functions are bounded. The above convergences will then lead to the convergence of the right-hand side of~(\ref{recursive-n}) to the right-hand side of~(\ref{recursive}) at an $s_{t1}$-independent rate. That is, $v_{nt}(s_{t1},\varepsilon(s^n_{t,-1}),x_{[t\bar t]},x_t)$ will converge to $v_t(s_{t1},\sigma_t,x_{[t\bar t]},x_t)$ at a rate independent of $s_{t1}$. We have completed the induction process. \qed

\noindent{\bf Proof of Theorem~\ref{main}: }Let us consider subgames starting with some time $t=1,2,...,\bar t$. For convenience, we let $\sigma_t=T_{[1,t-1]}(x^*_{[1,t-1]})\circ\sigma_1$. Now let $s_t=(s_{t1},s_{t2},...)$ be generated through sampling on $(S,{\cal B}(S),\sigma_t)$, $g=(g_1,g_2,...)$ be generated through sampling on $(G,{\cal B}(G),\gamma)$, and $i=(i_1,i_2,...)$ be generated through sampling on $(I,{\cal B}(I),\iota)$. In the remainder of the proof, we let $s^n_t=(s_{t1},...,s_{tn})$, $s^n_{t,-1}=(s_{t2},...,s_{tn})$, $g^n=(g_1,...,g_n)$, and $i^n=(i_1,...,i_n)$.

By Lemma~\ref{p-Prohorov} and Proposition~\ref{T-onestep}, we know that $\varepsilon(s^n_t)=\varepsilon(s_{t1},...,s_{tn})$ converges to $\sigma_t$ in probability, and also that $T_{nt}(x^*_t,g^n,i^n)\circ \varepsilon(s^n_t)$ converges to $T_t(x^*_t)\circ\sigma_t$ in probability. Due to Lemma~\ref{p-mc8}, $\varepsilon(s^n_{t,-1})$ and $[T_{nt}(x^*_t,g^n,i^n)\circ \varepsilon(s^n_t)]_{-1}$ will have the same respective convergences. Also, Lemma~\ref{joint} will lead to the convergence in probability of $\varepsilon(s^n_{t,-1},g^n_{-1})$ to $\sigma_t\times\gamma$. Due to $x_t$'s probabilistic continuity, Lemma~\ref{cojoint} will lead to the convergence in probability of $M_n(s^n_{t,-1},g^n_{-1},x_t)$ to $M(\sigma_t,x_t)$. Then,\\
\indent\M 1. $\psi_t(s_{t1},y(s_{t1},g_1),M_n(s^n_{t,-1},g_{-1},x_t))$ will converge to $\psi_t(s_{t1},y(s_{t1},g_1),M(\sigma_t,x_t))$ in probability at a $y$-independent rate due to (F2);\\
\indent\M 2. $v_{n,t+1}(\theta_t(s_{t1},y(s_{t1},g_1),M_n(s^n_{t,-1},g^n_{-1},x_t),i_1),[T_{nt}(x^*_t,g^n,i^n)\circ \varepsilon(s^n_t)]_{-1},x^*_{[t+1,\bar t]},x^*_{t+1})$ will converge to $v_{t+1}(\theta_t(s_{t1},y(s_{t1},g_1),M_n(s^n_{t,-1},g^n_{-1},x_t),i_1),T_t(x^*_t)\circ \sigma_t,x^*_{[t+1,\bar t]},x^*_{t+1})$ in probability at a $y$-independent rate due to Proposition~\ref{V-convergence}, which due to (S2) and Proposition~\ref{V-other}, will converge to $v_{t+1}(\theta_t(s_{t1},y(s_{t1},g_1),M(\sigma_t,x_t),i_1),T_t(x^*_t)\circ\sigma_t,x^*_{[t+1,\bar t]},x^*_{t+1})$ in probability at a $y$-independent rate.

As per-period payoffs are bounded, all value functions are bounded. By~(\ref{recursive}) and~(\ref{recursive-n}), the above convergences will then lead to the convergence of the left-hand side of~(\ref{second}) to the left-hand side of~(\ref{first}). At the same time, the right-hand side of~(\ref{second}) plus $\epsilon$ will converge to the right-hand side of~(\ref{first}) due to the convergence of $\varepsilon(s^n_{t,-1})$ to $\sigma_t$, Proposition~\ref{V-convergence}, and the uniform boundedness of the value functions. By~(\ref{first}), as long as $n$ is large enough,~(\ref{second}) will be true for any $\epsilon>0$ and $y\in {\cal M}(S\times G,X)$. This would then lead to the final conclusion due to Theorem~\ref{T-converge} and the boundedness of payoff functions.\qed

\section{Technical Developments in Section~\ref{stationary}}\label{app-c}

\noindent{\bf Proof of Theorem~\ref{main-s}: }Let $\epsilon>0$ be fixed. For $t=1,2,...$ satisfying $t\geq \ln(6\bar\psi/(\epsilon\cdot(1-\alpha)))/\ln(1/\alpha)+1$, we have from~(\ref{recursive-n-s}) and~(\ref{bounded-ss}),
\begin{equation}
\mid v_{n\infty}(s_1,\varepsilon(s_{-1}),x^*,y)-v_{nt}(s_1,\varepsilon(s_{-1}),x^*,y)\mid<\frac{\epsilon}{6}.
\end{equation}
Therefore, we need merely to select such a large $t$ and show that, when $n$ is large enough,
\begin{equation}\label{merely}
\int_{S^n} v_{n t}(s_1,\varepsilon(s_{-1}),x^*,x^*)\cdot (\sigma^*)^n(ds)\geq \int_{S^n} v_{n t}(s_1,\varepsilon(s_{-1}),x^*,y)\cdot (\sigma^*)^n(ds)-\frac{2\epsilon}{3}.
\end{equation}

For $t=1,2,...$, since $(x^*,\sigma^*)$ forms an equilibrium for $\Gamma$, we know~(\ref{first-s}) is true. This, as well as~(\ref{recursive-s}) and~(\ref{bounded-s}), lead to
\begin{equation}\label{first-s-tau}
\alpha^{t-\tau}\cdot\left[\int_S v_\tau(s,\sigma^*,x^*,y)\cdot \sigma^*(ds)-\int_S v_\tau(s,\sigma^*,x^*,x^*)\cdot \sigma^*(ds)\right]\leq \frac{2\alpha^{t-1}\cdot\bar\psi}{1-\alpha}\leq\frac{\epsilon}{3}.
\end{equation}
for $\tau=1,2,...,t$, $g\in G$, $s\in S$, and $y\in {\cal M}(S\times G,X)$.

We associate entities here with those defined in Section~\ref{games} when $\bar t$ there is fixed at the $t$ here. To signify the difference in the two notational systems, we add superscript ``$K$'' to symbols defined for the previous section. For instance, we write $v^K_\tau$ for the $v_\tau$ defined in that section, which has a different meaning than the $v_\tau$ here. Now, our $\alpha^{t-\tau}\cdot v_\tau(s,\sigma^*,x^*,y)$ can be understood as $v^K_{t+1-\tau}(s,\sigma^*,x',y)$, with $x'=(x'_{t+1-\tau},...,x'_t)\in ({\cal M}(S\times G,X))^\tau$ being such that $x'_{t'}=x^*$ for $t'=t+1-\tau,...,t$. Due to the association of $\sigma^*$ with $x^*$ through the definition~(\ref{compatible-s}), we can understand $\sigma^*$ as $T^K_{[1,\tau-1]}(x'_{[1,\tau-1]})\circ \sigma^K_1$, where $x'_{[1,\tau-1]}=(x'_1,...,x'_{\tau-1})\in ({\cal M}(S\times G,X))^{\tau-1}$ is such that $x'_{t'}=x^*$ for $t'=1,2,...,\tau-1$.

With these correspondences,~(\ref{first-s-tau}) can be translated into something akin to~(\ref{first}), with the only difference being that $-\epsilon/3$ should be added to all the right-hand sides. That is, we now know that the current $(x^*,\sigma^*)$ offers an $(\epsilon/3)$-Markov equilibrium for the nonatomic game $\Gamma^{\cal K}(\sigma^*)$ with $\bar t=t$, $\theta^K_\tau=\theta$, and $\psi^K_\tau=\alpha^{\tau-1}\cdot\psi$. Even though Theorem~\ref{main} is nominally about going from an 0-equilibrium for the nonatomic game to $\epsilon$-equilibria for finite games, we can follow exactly the same logic used to prove it to go from an $(\epsilon/3)$-equilibrium for the nonatomic game to $(2\epsilon/3)$-equilibria for finite games.

Thus, from one of the theorem's claims, we can conclude that, for $n$ large enough and any $y\in {\cal M}(S\times G,X)$,
\begin{equation}
\int_{S^n}\left(\sigma^K_1\right)^n(ds)\cdot v^K_{nt}\left(s_{1},\varepsilon(s_{-1}),x'_{[1t]},x'_1\right)\geq \int_{S^n}\left(\sigma^K_1\right)^n(ds)\cdot v^{K}_{nt}\left(s_{1},\varepsilon(s_{-1}),x'_{[1t]},y\right)-\frac{2\epsilon}{3},
\end{equation}
where $x'_{[1t]}$ is again to be understood as the policy that takes action $x^*(s,g)$ whenever the most immediate state-shock pair is $(s,g)$. But this translates into~(\ref{merely}). \qed

\end{document}